\documentclass[11pt]{article}

\usepackage{graphicx}
\usepackage{amsmath,amssymb,amsfonts,dsfont,slashed}
\usepackage{cite}
\usepackage{booktabs}

\allowdisplaybreaks[1]

\def\Xint#1{\mathchoice
   {\XXint\displaystyle\textstyle{#1}}%
   {\XXint\textstyle\scriptstyle{#1}}%
   {\XXint\scriptstyle\scriptscriptstyle{#1}}%
   {\XXint\scriptscriptstyle\scriptscriptstyle{#1}}%
   \!\int}
\def\XXint#1#2#3{{\setbox0=\hbox{$#1{#2#3}{\int}$}
     \vcenter{\hbox{$#2#3$}}\kern-0.5\wd0}}
\providecommand{\dashint}[1][0pt]{\Xint{\hspace{#1}-}}

\providecommand{\ec}{\;,}
\providecommand{\ep}{\;.}
\providecommand{\nt}{\notag}

\newcommand{\diff}{\text{d}}

\newcommand{\mpi}{M_{\pi}}
\newcommand{\mK}{M_K}
\newcommand{\meta}{M_\eta}

\newcommand{\muu}{m_\text{u}}
\newcommand{\md}{m_\text{d}}
\newcommand{\unity}{\mathds{1}}
\newcommand{\Fpi}{F_\pi}
\newcommand{\tpi}{t_\pi}
\newcommand{\tK}{t_K}
\newcommand{\tN}{t_N}
\newcommand{\tm}{t_\text{m}}

\newcommand{\ff}{\mathbf{f}}
\newcommand{\FF}{\mathbf{F}}
\newcommand{\GG}{\mathbf{G}}
\newcommand{\Del}{\boldsymbol{\Delta}}

\newcommand{\beq}{\begin{equation}}
\newcommand{\eeq}{\end{equation}}
\newcommand{\eps}{\epsilon}
\newcommand{\Order}{\mathcal{O}}

\renewcommand{\Im}{\text{Im}\,}
\renewcommand{\Re}{\text{Re}\,}

\providecommand{\unity}{\mathds{1}}

\providecommand{\GeV}{\,\text{GeV}}

\providecommand{\twosub}[1]{#1\big|^{2\text{-sub}}}

\begin{document}

\setlength{\unitlength}{1mm}

\renewcommand{\arraystretch}{1.2}

\numberwithin{equation}{section}

\author{M.~Hoferichter$^a$, C.~Ditsche$^a$, B.~Kubis$^a$, U.-G. Mei\ss ner$^{a,b}$}

\title{\bf Dispersive analysis of the scalar form factor\\ of the nucleon}

\date{}

\maketitle

\begin{center}
{\small
$^a${\it Helmholtz--Institut f\"ur Strahlen- und Kernphysik (Theorie) 
   and Bethe Center for Theoretical Physics, Universit\"at Bonn, D-53115 Bonn, Germany}

\bigskip

$^b${\it Institut f\"ur Kernphysik, Institute for Advanced Simulation, 
   and J\"ulich Center for Hadron Physics, Forschungszentrum J\"ulich, D-52425  J\"ulich, Germany}
}
\bigskip
\end{center}

\begin{abstract}
 Based on the recently proposed Roy--Steiner equations for pion--nucleon ($\pi N$) scattering~\cite{RS_DHKM}, we derive a system of coupled integral equations for the $\pi\pi\to\bar N N$ and $\bar K K\to\bar N N$ $S$-waves. These equations take the form of a two-channel Muskhelishvili--Omn\`es problem, whose solution in the presence of a finite matching point is discussed. We use these results to update the dispersive analysis of the scalar form factor of the nucleon fully including $\bar K K$ intermediate states. In particular, we determine the correction $\Delta_\sigma=\sigma(2\mpi^2)-\sigma_{\pi N}$, which is needed for the extraction of the pion--nucleon $\sigma$ term from $\pi N$ scattering, as a function of pion--nucleon subthreshold parameters and the $\pi N$ coupling constant. 
\end{abstract}

\section{Introduction}

The standard procedure to extract the pion--nucleon $\sigma$ term $\sigma_{\pi N}$, the nucleon form factor $\sigma(t)$ of the scalar current
$\hat m(\bar u u+\bar d d)$
at vanishing momentum transfer $t=0$ (here $\hat m=(\muu+\md)/2$ denotes the average mass of the light quarks), from $\pi N$ scattering data involves the venerable low-energy theorem that relates the Born-term-subtracted isoscalar $\pi N$ scattering amplitude at the Cheng--Dashen point to $\sigma(2\mpi^2)$~\cite{ChengDashen,BPP}. Later on, it was shown that the corrections to this low-energy theorem are very small, in particular they are free of chiral logarithms at full one-loop order in chiral perturbation theory (ChPT)~\cite{BKM96,BL01}. The extraction of $\sigma_{\pi N}$ itself thus requires knowledge of the difference
\beq
\label{Delta_sigma_def}
\Delta_\sigma=\sigma(2\mpi^2)-\sigma_{\pi N}\ec
\eeq
which can be determined by means of a dispersive representation of $\sigma(t)$~\cite{GLS90_FF}. The leading contribution to the imaginary part originates from $\pi\pi$ intermediate states, so that, upon neglecting higher terms in the spectral function, $\Im\sigma(t)$ can be expressed in terms of the scalar pion form factor $F_\pi^S(t)$ and the $\pi\pi\to\bar N N$ $S$-wave $f^0_+(t)$. Relying on the results of~\cite{DGL90} for $F_\pi^S(t)$ and of~\cite{Hoehler} for $f^0_+(t)$ led to the estimate~\cite{GLS90_FF} 
\beq
\label{GLSDelta}
\Delta_\sigma=(15.2\pm 0.4)\,{\rm MeV}\ec
\eeq
where the error only includes the uncertainty in the parameterization of the $\pi\pi$ phase available at that time. In particular, one should note that the contributions from $\bar K K$ intermediate states in the determination of $f^0_+(t)$ and the unitarity relation for $\sigma(t)$ were neglected, while being included in the calculation of $F_\pi^S(t)$. Although the dominant effect may indeed be expected in the pion form factor, such a treatment is strictly speaking inconsistent and leads to an additional uncertainty in~\eqref{GLSDelta} that is difficult to quantify. Moreover, the result for $f^0_+(t)$ from~\cite{Hoehler} corresponds to particular values of $\pi N$ subthreshold parameters and the $\pi N$ coupling constant. Especially the value $g^2/4\pi=14.28$ used for the coupling constant cannot be reconciled with more recent determinations~\cite{Nijmegen:1997,GWU:2006,piNcoupling:short,piNcoupling:long}.

In this article we perform an updated dispersive analysis of the scalar form factor of the nucleon. Based on Roy--Steiner (RS) equations for pion--nucleon scattering~\cite{RS_DHKM}, we determine the $\pi\pi\to\bar N N$ and $\bar K K\to\bar N N$ $S$-waves as solutions of a two-channel Muskhelishvili--Omn\`es (MO) problem and repeat the calculation of the scalar pion and kaon form factors for our input of $\pi\pi$ and $\pi\pi\to\bar K K$ partial waves. Taking everything together, we then analyze the spectral function of the scalar form factor of the nucleon fully including the effects from $\bar K K$ intermediate states, and provide an updated value for $\Delta_\sigma$ as a function of $\pi N$ subthreshold parameters and the $\pi N$ coupling constant.  

The paper is organized as follows. In Sect.~\ref{sec:MOsol_finite_tm} we consider a generic two-channel MO problem with finite matching point and develop a method to construct the corresponding Omn\`es matrix, which we apply to the coupled system of $\pi\pi\to\bar N N$ and $\bar K K\to\bar N N$ $S$-waves in Sect.~\ref{sec:pin_appl}. In Sect.~\ref{sec:SFF} we first present our results for the scalar pion and kaon form factors, which are then used as input for the dispersive analysis of the scalar form factor of the nucleon. We offer our conclusions in Sect.~\ref{sec:concl}. Several technical details of the calculation are relegated to the Appendices.

\section{Two-channel Muskhelishvili--Omn\`es problem}
\label{sec:MOsol_finite_tm}

We consider the generic coupled-channel integral equation
\beq
\ff(t)=\Del(t)+\frac{1}{\pi}\int\limits_{\tpi}^{\tm}\diff t'\frac{T^*(t')\Sigma(t')\ff(t')}{t'-t}
+\frac{1}{\pi}\int\limits_{\tm}^\infty\diff t'\frac{\Im \ff(t')}{t'-t}\ec
\eeq
where bold-faced quantities are two-dimensional vectors in channel space, representing pion and kaon intermediate states,\footnote{Throughout this work, we neglect $4\pi$ intermediate states, which are phenomenologically irrelevant at low energies. In fact, the effective onset of $4\pi$ inelasticities marks the breakdown of the present two-channel model, cf.\ Sect.~\ref{sec:pin_appl}.} and the imaginary part of $\ff(t)$ is assumed to be known above the matching point $\tm$. In particular, $f_1(t)$ and $f_2(t)$ can be thought of as the $\pi\pi\to\bar N N$ and $\bar K K\to\bar N N$ $S$-waves, respectively, although the following discussion can be carried out for general angular momentum $J$. The function $\Del(t)$ contains at most left-hand cuts and is therefore real for $t\geq \tpi=4\mpi^2$.
The unitarity relation is written in the form
\beq
 \Im \ff(t)=T^*(t)\Sigma(t)\ff(t)\ec
\eeq
with $T$-matrix $T(t)$ and phase-space factor $\Sigma(t)$ parameterized as\footnote{The masses of nucleon, pion, and kaon are denoted by $m$, $\mpi$, and $\mK$, respectively, and defined by the charged-particle masses as given in~\cite{PDG}.} 
\beq
\label{Tmatrix}
T(t)=\begin{pmatrix}
      \frac{\eta(t) e^{2i\delta(t)}-1}{2i\sigma_t^\pi q_t^{2J}} & |g(t)|e^{i\psi(t)}\\ 
  |g(t)|e^{i\psi(t)} & \frac{\eta(t) e^{2i(\psi(t)-\delta(t))}-1}{2i\sigma_t^K k_t^{2J}}
\end{pmatrix}\ec\qquad
\Sigma(t)=\text{diag}\Big(\sigma_t^\pi q_t^{2J}\theta\big(t-\tpi\big), \sigma_t^Kk_t^{2J}\theta\big(t-\tK\big)\Big)\ec
\eeq
with pion, kaon, and---for later use---nucleon $t$-channel momenta
\beq
q_t=\sqrt{\frac{t}{4}-\mpi^2}=\frac{\sqrt{t}}{2}\sigma_t^\pi\ec\qquad k_t=\sqrt{\frac{t}{4}-\mK^2}=\frac{\sqrt{t}}{2}\sigma_t^K\ec\qquad
p_t=\sqrt{\frac{t}{4}-m^2}=\frac{\sqrt{t}}{2}\sigma_t^N \ec
\eeq
and the two-kaon threshold $\tK=4\mK^2$.
The scattering phases $\delta(t)$ and $\psi(t)$ are required as input for $\tpi\leq t\leq\tm$, where Watson's theorem~\cite{Watson} demands $\psi(t)=\delta(t)$ for $t\leq \tK$. Moreover, the modulus of the $\pi\pi\to\bar K K$ $S$-wave $g(t)$ is needed in the full range $\tpi\leq t\leq\tm$, and thus has to be analytically continued into the pseudophysical region $\tpi\leq t\leq \tK$. Finally, the inleasticity parameter $\eta(t)$ can be related to $|g(t)|$ via    
\beq
\eta(t)=\sqrt{1-4\sigma_t^\pi\sigma_t^K(q_tk_t)^{2J}|g(t)|^2\theta\big(t-\tK\big)}\ec
\eeq
and the relation between $S$- and $T$-matrix reads
\beq
S(t)=\unity+2i\,\Sigma^{1/2}(t)T(t)\Sigma^{1/2}(t)\ep
\eeq

\subsection{Formal solution}

We define the Omn\`es matrix $\Omega(t)$ by
\beq
\label{def_finite}
\begin{Bmatrix}\Im \Omega(t)=T^*(t)\Sigma(t)\Omega(t) \\ \Im \Omega(t)=0\end{Bmatrix}
\qquad \text{for} \qquad
\begin{Bmatrix}\tpi\leq t \leq \tm \\ \text{otherwise}\end{Bmatrix}
\eeq
and choose the normalization $\Omega(0)=\unity$. Writing
\beq
\FF(t)=\ff(t)-\Del(t)=\Omega(t)\GG(t)\ec
\eeq
it follows that for $t\geq \tpi$ 
\beq
\big(\unity-2i\, T^*(t)\Sigma(t)\big)\Omega(t_+)\big(\GG(t_+)-\GG(t_-)\big)=2i\,T^*(t)\Sigma(t)\Del(t)\ec
\eeq
where $t_\pm=t\pm i\epsilon$ and the physical limit is given by $t_+$.
Using unitarity in the form
\beq
\big(\unity-2i\, T^*(t)\Sigma(t)\big)^{-1}=\unity+2i\,T(t)\Sigma(t)\ec
\eeq
which in particular holds for $t\leq \tK$ by virtue of Watson's theorem,
we find
\beq
\GG(t_+)-\GG(t_-)=2i\,\Omega^{-1}(t)T(t)\Sigma(t)\Del(t)\ec
\eeq
and thus
\begin{align}
\label{omnes_fin}
\ff(t)&=\Del(t)+\frac{\Omega(t)}{\pi}\int\limits_{\tpi}^{\tm}\diff t'\frac{\Omega^{-1}(t')T(t')\Sigma(t')\Del(t')}{t'-t}+
\frac{\Omega(t)}{\pi}\int\limits_{\tm}^\infty\diff t'\frac{\Omega^{-1}(t')\Im \ff(t')}{t'-t}\nt\\
&=\Del(t)-\frac{\Omega(t)}{\pi}\int\limits_{\tpi}^{\tm}\diff t'\frac{\Im\Omega^{-1}(t')\Del(t')}{t'-t}+
\frac{\Omega(t)}{\pi}\int\limits_{\tm}^\infty\diff t'\frac{\Omega^{-1}(t')\Im \ff(t')}{t'-t}\ep
\end{align}
The problem is such reduced to finding a matrix $\Omega(t)$ that fulfills~\eqref{def_finite}. For $\tpi\leq t\leq \tK$ we have
\beq
\Omega(t_+)=\big(\unity+2i\,T(t)\Sigma(t)\big)\Omega(t_-)\ep
\eeq
Taking the determinant on both sides yields, again using Watson's theorem for $t\leq \tK$,
\beq
\det \Omega(t_+)=e^{2i\psi(t)}\det\Omega(t_-)\ec
\eeq
and thus~\cite{Moussallam99}
\beq
\label{det_fin}
\det \Omega(t)=\exp\Bigg\{\frac{t}{\pi}\int\limits_{\tpi}^{\tm}\diff t'\frac{\psi(t')}{t'(t'-t)}\Bigg\}\ep
\eeq
Although the determinant allows for an analytic solution in the same way as in the single-channel case~\cite{Omnes}, there is in general no analytic solution for the Omn\`es matrix itself even for an infinite matching point, which therefore has to be calculated numerically, either by an iterative procedure~\cite{DGL90} or a discretization method, i.e.\ solving a matrix equation~\cite{Moussallam99,SDG_diss} (for a mathematician's point of view see~\cite{Muskhelishvili}). Similarly to the single-channel case, we expect a cusp at $\tm$, which has to be taken into account in the numerical evaluation of the integrals in~\eqref{omnes_fin} (see~\cite{BDM04,RS_DHKM}). Indeed, for $t\to\tm$ the determinant behaves as
\beq
\label{det_asym}
\det\Omega(t)\sim |\tm-t|^x\ec\qquad x=\frac{\psi(\tm)}{\pi}\ep
\eeq
Accordingly, we write for $t\to\tm$ from below
\beq
\label{def_below}
 \det\Omega(t)=\det\bar\Omega(\tm)e^{i\pi x} |\tm-t|^x\ec\qquad \Omega_{ij}(t)=\bar\Omega_{ij}(\tm)e^{i\delta_{ij}(\tm)}|\tm-t|^{x_{ij}}\ec
\eeq
and from above
\beq
\label{def_above}
 \det\Omega(t)=\det\bar\Omega(\tm) |\tm-t|^x\ec\qquad \Omega_{ij}(t)=\bar\Omega_{ij}(\tm)|\tm-t|^{x_{ij}}\ec
\eeq
since $\Omega(t)$ is real above $\tm$. Here, we have assumed that the (real) functions $\bar\Omega_{ij}(t)$ are continuous at $\tm$. The strength of the cusp in each component $\Omega_{ij}(t)$ of the Omn\`es matrix is determined by the numbers $x_{ij}$, whose relation to the $S$-matrix parameters will be established in the following sections. Throughout this paper we will consider the case $0< x_{ij} < 1$, which is relevant for the coupled-channel $S$-wave system of $\pi\pi$ and $\bar K K$ intermediate states. The extension to arbitrary values of $x_{ij}$ can then be done along the lines described in~\cite{BDM04,HPS}.

\subsection{Dispersive representation of the Omn\`es matrix}

For $0<x_{ij}<1$ we may write down a dispersive representation
\beq
\Omega(t)=\Big(1-\frac{t}{\tm}\Big)\unity+\frac{t(t-\tm)}{\pi}\int\limits_{\tpi}^{\tm}\diff t'\frac{T^*(t')\Sigma(t')\Omega(t')}{t'(t'-\tm)(t'-t)}\ec
\eeq
where the subtraction constants have been fixed in such a way that $\Omega(0)=\unity$ and $\Omega(\tm)=0$.
In particular, we can investigate the limit $t\to\tm$ to obtain information on $x_{ij}$. Using the asymptotic form of the integrals~\cite{BDM04} (the dash denotes the principal-value part of the integral)
\begin{align}
\label{integrals}
&\frac{|\tm-t|^{x}}{\pi}\dashint[9pt]\limits_{\tm-\eps}^{\tm}\frac{\diff t'}{(t'-t)|\tm-t'|^x}
\overset{\eps\to 0,\,t\to\tm}{\longrightarrow}\frac{1}{\pi}\dashint[0.5pt]\limits_0^\infty\frac{\diff v}{v^x(1-v)}=-\cot\pi x\ec\nt\\
&\frac{|\tm-t|^{x}}{\pi}\int\limits_{\tm}^{\tm+\eps}\frac{\diff t'}{(t'-t)|\tm-t'|^x}
\overset{\eps\to 0,\,t\to\tm}{\longrightarrow}\frac{1}{\pi}\int\limits_0^\infty\frac{\diff v}{v^x(1+v)}=\frac{1}{\sin \pi x}\ec
\end{align}
we obtain for $t\to\tm$ from below\footnote{We consider the case $J=0$ for simplicity. The general case can always be recovered by introducing the correct phase-space factors according to $\sigma_t^\pi\to \sigma_t^\pi q_t^{2J}$ and $\sigma_t^K\to \sigma_t^K k_t^{2J}$.}
{\allowdisplaybreaks
\begin{align}
\label{Omegaij_above}
 \bar\Omega_{11}|\tm-t|^{x_{11}}e^{i\delta_{11}}&=\bar\Omega_{11}|\tm-t|^{x_{11}}e^{i\delta_{11}}\frac{e^{i\pi x_{11}}}{\sin\pi x_{11}}\frac{1-\eta e^{-2i\delta}}{2i}+\bar\Omega_{21}|\tm-t|^{x_{21}}e^{i\delta_{21}}\frac{e^{i\pi x_{21}}}{\sin\pi x_{21}}|g|\sigma_{\tm}^Ke^{-i\psi}\ec\nt\\
 \bar\Omega_{12}|\tm-t|^{x_{12}}e^{i\delta_{12}}&=\bar\Omega_{12}|\tm-t|^{x_{12}}e^{i\delta_{12}}\frac{e^{i\pi x_{12}}}{\sin\pi x_{12}}\frac{1-\eta e^{-2i\delta}}{2i}+\bar\Omega_{22}|\tm-t|^{x_{22}}e^{i\delta_{22}}\frac{e^{i\pi x_{22}}}{\sin\pi x_{22}}|g|\sigma_{\tm}^Ke^{-i\psi}\ec\nt\\
\bar\Omega_{21}|\tm-t|^{x_{21}}e^{i\delta_{21}}&=\bar\Omega_{11}|\tm-t|^{x_{11}}e^{i\delta_{11}}\frac{e^{i\pi x_{11}}}{\sin\pi x_{11}}|g|\sigma_{\tm}^\pi e^{-i\psi}\nt\\
&\qquad+\bar\Omega_{21}|\tm-t|^{x_{21}}e^{i\delta_{21}}\frac{e^{i\pi x_{21}}}{\sin\pi x_{21}}\frac{1-\eta e^{-2i(\psi-\delta)}}{2i}\ec\nt\\
\bar\Omega_{22}|\tm-t|^{x_{22}}e^{i\delta_{22}}&=\bar\Omega_{12}|\tm-t|^{x_{12}}e^{i\delta_{12}}\frac{e^{i\pi x_{12}}}{\sin\pi x_{12}}|g|\sigma_{\tm}^\pi e^{-i\psi}\nt\\
&\qquad+\bar\Omega_{22}|\tm-t|^{x_{22}}e^{i\delta_{22}}\frac{e^{i\pi x_{22}}}{\sin\pi x_{22}}\frac{1-\eta e^{-2i(\psi-\delta)}}{2i}\ec
\end{align}
and for $t\to\tm$ from above
\begin{align}
\label{Omegaij_below}
 \bar\Omega_{11}|\tm-t|^{x_{11}}&=\bar\Omega_{11}|\tm-t|^{x_{11}}\frac{e^{i\delta_{11}}}{\sin\pi x_{11}}\frac{1-\eta e^{-2i\delta}}{2i}+\bar\Omega_{21}|\tm-t|^{x_{21}}\frac{e^{i\delta_{21}}}{\sin\pi x_{21}}|g|\sigma_{\tm}^K e^{-i\psi}\ec\nt\\
 \bar\Omega_{12}|\tm-t|^{x_{12}}&=\bar\Omega_{12}|\tm-t|^{x_{12}}\frac{e^{i\delta_{12}}}{\sin\pi x_{12}}\frac{1-\eta e^{-2i\delta}}{2i}+\bar\Omega_{22}|\tm-t|^{x_{22}}\frac{e^{i\delta_{22}}}{\sin\pi x_{22}}|g|\sigma_{\tm}^K e^{-i\psi}\ec\nt\\
\bar\Omega_{21}|\tm-t|^{x_{21}}&=\bar\Omega_{11}|\tm-t|^{x_{11}}\frac{e^{i\delta_{11}}}{\sin\pi x_{11}}|g|\sigma_{\tm}^\pi e^{-i\psi}
+\bar\Omega_{21}|\tm-t|^{x_{21}}\frac{e^{i\delta_{21}}}{\sin\pi x_{21}}\frac{1-\eta e^{-2i(\psi-\delta)}}{2i}\ec\nt\\
\bar\Omega_{22}|\tm-t|^{x_{22}}&=\bar\Omega_{12}|\tm-t|^{x_{12}}\frac{e^{i\delta_{12}}}{\sin\pi x_{12}}|g|\sigma_{\tm}^\pi e^{-i\psi}
+\bar\Omega_{22}|\tm-t|^{x_{22}}\frac{e^{i\delta_{22}}}{\sin\pi x_{22}}\frac{1-\eta e^{-2i(\psi-\delta)}}{2i}\ec
\end{align}
}\noindent
where we have suppressed the evaluation at $\tm$ wherever possible. If we assume that $g(\tm)\neq 0$ and $\bar \Omega_{ij}(\tm)\neq 0$ (which can always be achieved by choosing the matching point appropriately), we can conclude from the first line of~\eqref{Omegaij_above} that $x_{21}\geq x_{11}$, since otherwise $g$ or $\bar\Omega_{21}$ would have to vanish at $\tm$. Conversely, the third line requires $x_{21}\leq x_{11}$ by the same argument, and hence $x_{11}=x_{21}$. Similarly, we find $x_{12}=x_{22}$. Moreover, as the determinant behaves according to~\eqref{det_asym}, we can conclude that $x_{11}+x_{12}=x$, again provided that $\det\bar\Omega (\tm)\neq0$. Dividing the first line of~\eqref{Omegaij_above} by the first line of~\eqref{Omegaij_below}, we find 
\beq
e^{i\delta_{11}}\bigg(1-\frac{e^{i\pi x_{11}}}{\sin\pi x_{11}}\frac{1-\eta e^{-2i\delta}}{2i}\bigg)
=e^{i\pi x_{21}}\bigg(1-\frac{e^{i\delta_{11}}}{\sin\pi x_{11}}\frac{1-\eta e^{-2i\delta}}{2i}\bigg)\ec
\eeq
which for $x_{21}=x_{11}$ reduces to
\beq
e^{i\delta_{11}}=e^{i\pi x_{11}}\ec
\eeq
and thus $\pi x_{11}=\delta_{11}$ up to integer multiples of $2\pi$.  
Arguing analogously for $x_{12}$, these results can be summarized as
\beq
\label{xij}
x_{11}=x_{21}\ec\qquad x_{12}=x_{22}\ec\qquad x_{ij}=\frac{\delta_{ij}}{\pi}\ec \qquad x_{11}+x_{12}=x\ep
\eeq
By virtue of~\eqref{xij}, \eqref{Omegaij_above} and~\eqref{Omegaij_below} take the form
\begin{align}
\label{spectral_rep_const}
\renewcommand{\arraystretch}{1.3}
 \begin{pmatrix}
 \frac{e^{i\pi x_{11}}}{\sin\pi x_{11}}\frac{1-\eta e^{-2i\delta}}{2i}-1 & \frac{e^{i\pi x_{11}}}{\sin\pi x_{11}}|g|\sigma_{\tm}^K e^{-i\psi}\\
\frac{e^{i\pi x_{11}}}{\sin\pi x_{11}}|g|\sigma_{\tm}^\pi e^{-i\psi} & \frac{e^{i\pi x_{11}}}{\sin\pi x_{11}}\frac{1-\eta e^{-2i(\psi-\delta)}}{2i}-1
\end{pmatrix}
\begin{pmatrix}
 \bar\Omega_{11}\\ 
\bar\Omega_{21}
\end{pmatrix}
&=0\ec\nt\\
\renewcommand{\arraystretch}{1.3}
 \begin{pmatrix}
 \frac{e^{i\pi x_{12}}}{\sin\pi x_{12}}\frac{1-\eta e^{-2i\delta}}{2i}-1 & \frac{e^{i\pi x_{12}}}{\sin\pi x_{12}}|g|\sigma_{\tm}^K e^{-i\psi}\\
\frac{e^{i\pi x_{12}}}{\sin\pi x_{12}}|g|\sigma_{\tm}^\pi e^{-i\psi} & \frac{e^{i\pi x_{12}}}{\sin\pi x_{12}}\frac{1-\eta e^{-2i(\psi-\delta)}}{2i}-1
\end{pmatrix}
\begin{pmatrix}
 \bar\Omega_{12}\\ 
\bar\Omega_{22}
\end{pmatrix}
&=0\ep
\end{align}
\noindent
To ensure the existence of non-trivial solutions the determinants of the coefficient matrices must vanish. This leads to
\beq
\cos \pi (2x_{11}-x)-\eta\cos\pi (2y-x)=0\ec\qquad \cos \pi (2x_{12}-x)-\eta\cos\pi (2y-x)=0\ec\qquad y=\frac{\delta(\tm)}{\pi}\ep
\eeq
These conditions are invariant under $x_{11}\to x-x_{11}$, i.e.~there is an ambiguity between $x_{11}$ and $x_{12}$. However, demanding that $x_{11}$ coincide with $x$ in the single-channel limit yields
\beq
\label{x11result}
x_{11}=\frac{1}{2}\Big\{x+\frac{1}{\pi}\arccos(\eta\cos\pi(2y-x))\Big\}\ec\qquad x_{12}=\frac{1}{2}\Big\{x-\frac{1}{\pi}\arccos(\eta\cos\pi(2y-x))\Big\}\ep
\eeq
Finally, \eqref{spectral_rep_const} together with~\eqref{x11result} and
\begin{align}
\label{sinx11}
 \sin\pi x_{11}&=\sin\frac{\pi x}{2}\sqrt{\frac{1+z}{2}}+\cos\frac{\pi x}{2}\sqrt{\frac{1-z}{2}}\ec\qquad
 \cos\pi x_{11}=\cos\frac{\pi x}{2}\sqrt{\frac{1+z}{2}}-\sin\frac{\pi x}{2}\sqrt{\frac{1-z}{2}}\ec\nt\\
\sin\pi x_{12}&=\sin\frac{\pi x}{2}\sqrt{\frac{1+z}{2}}-\cos\frac{\pi x}{2}\sqrt{\frac{1-z}{2}}\ec\qquad
 \cos\pi x_{12}=\cos\frac{\pi x}{2}\sqrt{\frac{1+z}{2}}+\sin\frac{\pi x}{2}\sqrt{\frac{1-z}{2}}\ec\nt\\
z&=\eta\cos\pi(2y-x)\ec
\end{align}
 can be used to derive constraints on $\bar\Omega_{ij}$. We find
\beq
\label{ratio_Omij}
 \frac{\bar\Omega_{21}}{\bar\Omega_{11}}=\frac{N}{2|g|\sigma_{\tm}^K}\ec\qquad
 \frac{\bar\Omega_{12}}{\bar\Omega_{22}}=-\frac{N}{2|g|\sigma_{\tm}^\pi}\ec\qquad N=\sqrt{1-\eta^2\cos^2\pi(2y-x)}-\eta\sin\pi(2y-x)\ep
\eeq
In the single-channel case one can show that, using the integrals~\eqref{integrals},
the solution for $f(t)$ is automatically continuous at $\tm$~\cite{BDM04}. In fact, the same statement holds true also in the two-channel case. The relations~\eqref{ratio_Omij} are essential in the proof, as demonstrated in Appendix~\ref{app:continuity}.

\subsection{Construction of the Omn\`es matrix}

\subsubsection{Infinite matching point}

Our construction of the two-channel Omn\`es matrix with finite matching point will heavily rely on the solution for its infinite-matching-point analog
 $\Omega^\infty(t)$, whose defining property can be stated as
\beq
\label{Omnes_def}
\begin{Bmatrix}\Im \Omega^\infty(t)=T^*(t)\Sigma(t)\Omega^\infty(t) \\ \Im \Omega^\infty(t)=0\end{Bmatrix}
\qquad \text{for} \qquad
\begin{Bmatrix}t\geq \tpi \\ \text{otherwise}\end{Bmatrix}\ep
\eeq
For its calculation we follow~\cite{Moussallam99} and discretize the unsubtracted dispersion relation
\beq
\Re \Omega^\infty(t)=\frac{1}{\pi}\dashint[0.5pt]\limits_{\tpi}^{\infty}\diff t'\frac{\Im \Omega^\infty(t')}{t'-t}
\eeq
on a set of Gau\ss--Legendre integration points. Note that an unsubtracted dispersion relation converges provided that the phase-shift at infinity is positive. In the one-channel case this can be directly deduced from the explicit solution
\beq
\label{sol_single_infinite}
\Omega^\infty(t)=\exp\Bigg\{\frac{t}{\pi}\int\limits_{\tpi}^\infty\diff t'\frac{\delta(t')}{t'(t'-t)}\Bigg\}\ec
\eeq
which behaves as 
\beq
\Omega^\infty(t)\sim t^{-\frac{\delta(\infty)}{\pi}}
\eeq
for large $t$. 
The unitarity condition~\eqref{Omnes_def} can be rewritten as
\beq
\label{unitarity}
\renewcommand{\arraystretch}{1.3}
\Im\mathbf{\Omega}^\infty_i=\begin{pmatrix}
                    \frac{\eta\sin(2\delta-\psi)+\sin\psi}{\eta\cos(2\delta-\psi)+\cos\psi} & \frac{2|g|\sigma_t^K\theta(t-\tK)}{\eta\cos(2\delta-\psi)+\cos\psi}\\
\frac{2|g|\sigma_t^\pi}{\eta\cos(2\delta-\psi)+\cos\psi} & -\frac{\eta \sin(2\delta-\psi)-\sin\psi}{\eta\cos(2\delta-\psi)+\cos\psi}
                   \end{pmatrix}
\Re\mathbf{\Omega}^\infty_i\ec\qquad i\in\{1,2\}\ec
\eeq
with
\beq
\label{Omnes_vector}
\mathbf{\Omega}_1^\infty=\begin{pmatrix}
                                          \Omega^\infty_{11}\\ \Omega^\infty_{21}
                                         \end{pmatrix}\ec \qquad
\mathbf{\Omega}_2^\infty=\begin{pmatrix}
                                          \Omega^\infty_{12}\\ \Omega^\infty_{22}
                                         \end{pmatrix}\ec
\eeq
which below the two-kaon threshold reduces to
\beq
\label{Omnes_below_tK}
\Im\mathbf{\Omega}^\infty_i=\begin{pmatrix}
                    \tan\delta & 0\\
\frac{|g|\sigma_t^\pi}{\cos\delta} & 0
                   \end{pmatrix}
\Re\mathbf{\Omega}^\infty_i\ep
\eeq
The details of the numerical solution of the corresponding integral equation for $\Re\mathbf{\Omega}^\infty_i$ are described in~\cite{Moussallam99,SDG_diss}.

\subsubsection{Finite matching point, single-channel case}
\label{sec:fmp_single}

An Omn\`es function with a finite matching point does not allow for an unsubtracted dispersion relation, since the solution~\eqref{det_fin}
tends to a constant for $t\to\infty$, and consequently one picks up contributions at infinity. Moreover, the cusp at the matching point renders both the discretization method and an iterative procedure involving subtracted dispersion relations inappropriate, as neither is able to accurately reproduce the analytic behavior around $\tm$. For this reason, the aim of this section is to establish a method that relies on the known solution in the infinite-matching-point scenario.

We first observe that the function
\beq
\xi(t)=\bigg(\frac{\tm-t}{\tm}\bigg)^{x(t)}=\bigg|\frac{\tm-t}{\tm}\bigg|^{x(t)}e^{-i\pi x(t)\theta(t-\tm)}
\eeq
has the correct properties to cancel an imaginary part above $\tm$. Indeed, the function
\beq
\label{Omnes_constr}
\Omega(t)=\Omega^\infty(t)\xi(t)\ec
\eeq
with $\Omega^\infty(t)$ from~\eqref{sol_single_infinite}, fulfills
\beq
\begin{Bmatrix}\Im \Omega(t)=T^*(t)\Sigma(t)\Omega(t) \\ 
\Im \Omega(t)=\Omega^\infty(t)|\xi(t)|\Big(e^{i(\pi x(t)-\delta(t))}\sin\delta(t)-\sin\pi x(t)\Big)\end{Bmatrix}
\qquad \text{for} \qquad
\begin{Bmatrix}\tpi\leq t \leq \tm \\ t\geq\tm\end{Bmatrix}\ec
\eeq
and with the choice
\beq
\begin{Bmatrix}x(t)=\frac{\delta(\tm)}{\pi} \\ 
x(t)=\frac{\delta(t)}{\pi}\end{Bmatrix}
\qquad \text{for} \qquad
\begin{Bmatrix}t \leq \tm \\ t\geq\tm\end{Bmatrix}
\eeq
the defining property~\eqref{def_finite} holds. Since we know the analytic solution, we can study the properties of this construction in more detail
\beq
\label{omnes_sol}
\Omega(t)=\exp\Bigg\{\frac{t}{\pi}\int\limits_{\tpi}^{\tm}\diff t'\frac{\delta(t')}{t'(t'-t)}\Bigg\}
\exp\Bigg\{\frac{t}{\pi}\int\limits_{\tm}^{\infty}\diff t'\frac{\delta(t')-\pi x(t)}{t'(t'-t)}\Bigg\}\ep
\eeq
The first term coincides with the expected result for the single-channel case, while the second factor is new. It is indeed real, and thus preserves all defining properties, in particular the normalization $\Omega(0)=1$. This example shows that we would exactly recover the result~\eqref{det_fin} if we did not know the solution for a finite matching point, constructed it according to~\eqref{Omnes_constr}, and chose $\delta(t)=\delta(\tm)$ for $t$ above the matching point. Obviously, \eqref{omnes_sol} implies that ``the'' Omn\`es function is not unique. The derivation of~\eqref{omnes_fin}, however, only relies on the defining properties and is therefore independent of such modifications.

\subsubsection{Finite matching point, two-channel case}

To generalize the preceding considerations to the two-channel case we define a matrix
\beq
\xi_{ij}(t)=\bar \xi_{ij}(t)  \bigg(\frac{\tm-t}{\tm}\bigg)^{x_{ij}(t)}
\eeq
with real functions $\bar\xi_{ij}(t)$. In view of the results of the previous section we take phases and inelasticities constant above $\tm$ and thus can ignore the $t$-dependence of $x_{ij}$ and $\bar\xi_{ij}$, which in the following will always be understood to be evaluated at $\tm$. It is straightforward to show that 
\beq
\label{sol_fin_Omega_til}
\begin{Bmatrix}\Omega(t)=a(t)\Omega^\infty(t) \xi(t) \\ 
\Omega(t)=a(t)(\Omega^\infty)^T(t)\xi(t)\end{Bmatrix}
\qquad \text{for} \qquad
\begin{Bmatrix}t \leq \tm \\ t\geq\tm\end{Bmatrix}\ec
\eeq
with infinite-matching-point solution $\Omega^\infty(t)$ and a real matrix $a(t)$, fulfills~\eqref{def_finite} provided that
\begin{align}
\label{cond_xi}
\Im\xi^T(t)+\xi^T(t) T(t)\Sigma(t)&=0\qquad \text{for}\ t\geq\tm\ec\nt\\
\big[a(t),T^*(t)\Sigma(t)\big]&=0\qquad \text{for}\ t\leq\tm\ep
\end{align} 
Imposing $x_{11}=x_{21}=x-x_{12}=x-x_{22}$, the first condition corresponds to
\begin{align}
\label{eq_xi}
\renewcommand{\arraystretch}{1.3}
 \begin{pmatrix}
 \frac{e^{i\pi x_{11}}}{\sin\pi x_{11}}\frac{1-\eta e^{-2i\delta}}{2i}-1 & \frac{e^{i\pi x_{11}}}{\sin\pi x_{11}}|g|\sigma_{\tm}^\pi e^{-i\psi}\\
\frac{e^{i\pi x_{11}}}{\sin\pi x_{11}}|g|\sigma_{\tm}^K e^{-i\psi} & \frac{e^{i\pi x_{11}}}{\sin\pi x_{11}}\frac{1-\eta e^{-2i(\psi-\delta)}}{2i}-1
\end{pmatrix}
\begin{pmatrix}
 \bar\xi_{11}\\ 
\bar\xi_{21}
\end{pmatrix}
&=0\ec\nt\\
\renewcommand{\arraystretch}{1.3}
 \begin{pmatrix}
 \frac{e^{i\pi x_{12}}}{\sin\pi x_{12}}\frac{1-\eta e^{-2i\delta}}{2i}-1 & \frac{e^{i\pi x_{12}}}{\sin\pi x_{12}}|g|\sigma_{\tm}^\pi e^{-i\psi}\\
\frac{e^{i\pi x_{12}}}{\sin\pi x_{12}}|g|\sigma_{\tm}^K e^{-i\psi} & \frac{e^{i\pi x_{12}}}{\sin\pi x_{12}}\frac{1-\eta e^{-2i(\psi-\delta)}}{2i}-1
\end{pmatrix}
\begin{pmatrix}
 \bar\xi_{12}\\ 
\bar\xi_{22}
\end{pmatrix}
&=0\ep
\end{align}
Non-trivial solutions of~\eqref{eq_xi} again exist for $x_{11}$ and $x_{12}$ given by~\eqref{x11result}, while the components of $\xi(t)$ are related by
\beq
 \frac{\bar\xi_{21}}{\bar\xi_{11}}=\frac{N}{2|g|\sigma_{\tm}^\pi}\ec\qquad \frac{\bar\xi_{12}}{\bar\xi_{22}}=-\frac{N}{2|g|\sigma_{\tm}^K}\ep
\eeq
For definiteness, we take
\beq
\bar\xi_{11}=1\ec\qquad \bar\xi_{22}=\bigg(1+\frac{N^2}{1-\eta^2}\bigg)^{-1}\ec
\eeq
which ensures that
\beq
\det\xi(t)=\bigg(\frac{\tm-t}{\tm}\bigg)^x\ec
\eeq
and thus, by the results of the previous section, would preserve the form~\eqref{det_fin} of the determinant of the Omn\`es function if $\det a(t)=1$. 
The condition~\eqref{cond_xi} on $a(t)$ requires
\beq
a_{21}=a_{12}\frac{\sigma_t^\pi}{\sigma_t^K}\ec\qquad a_{22}=a_{11}-a_{12}\frac{\eta\sin(2\delta-\psi)}{|g|\sigma_t^K}\ec
\eeq
while $a_{11}$ and $a_{12}$ can be chosen freely.

Finally, $\Omega(t)$ should fulfill the normalization $\Omega(0)=\mathds{1}$, which can be achieved by modifying the normalization in the calculation of $\Omega^\infty(t)$ appropriately. Assuming $a(0)=\mathds{1}$, the corresponding condition 
\beq
\Omega^\infty(0)\xi(0)=\begin{pmatrix}
                 \bar\xi_{11}\big(\Omega^\infty_{11}(0)+\frac{N}{2|g|\sigma_{\tm}^\pi}\Omega^\infty_{12}(0)\big) & \bar\xi_{22}\big(\Omega^\infty_{12}(0)-\frac{N}{2|g|\sigma_{\tm}^K}\Omega^\infty_{11}(0)\big)\\
		  \bar\xi_{11}\big(\Omega^\infty_{21}(0)+\frac{N}{2|g|\sigma_{\tm}^\pi}\Omega^\infty_{22}(0)\big) & \bar\xi_{22}\big(\Omega^\infty_{22}(0)-\frac{N}{2|g|\sigma_{\tm}^K}\Omega^\infty_{21}(0)\big)
                \end{pmatrix}
=\mathds{1}
\eeq
leads to
\beq
\Omega^\infty(0)=\bigg(1+\frac{N^2}{1-\eta^2}\bigg)^{-1}\begin{pmatrix}
                 {\bar\xi_{11}}^{-1} & \bar\xi_{11}^{-1}\frac{N}{2|g|\sigma_{\tm}^K}\\
		  -\bar\xi_{22}^{-1}\frac{N}{2|g|\sigma_{\tm}^\pi} & \bar\xi_{22}^{-1}
                \end{pmatrix}\ep
\eeq
The above construction~\eqref{sol_fin_Omega_til} ensures that 
$\Omega_{ij}(t)$ has the expected behavior for $t\to\tm$,
that the factors of $(1-t/\tm)^{x_{ij}}$ factorize in~\eqref{eq_xi} (which have therefore already been canceled there),
and that $\det\bar \Omega(t)$ is continuous at $\tm$. However, 
 in general, $\bar \Omega_{ij}(t)$ itself will not be continuous at $\tm$
and the condition~\eqref{ratio_Omij} will be violated.

In order to remove these shortcomings we make use of the freedom in choosing $a(t)$. In fact, a particular choice of $a(t)$ can  enforce continuity of either $\bar{\mathbf{\Omega}}_1(t)$ or $\bar{\mathbf{\Omega}}_2(t)$, but not of both simultaneously. This impediment can be circumvented by noting that~\eqref{sol_fin_Omega_til} may be regarded as separate equations for $\bar{\mathbf{\Omega}}_1(t)$ and $\bar{\mathbf{\Omega}}_2(t)$. We can thus derive an $\bar{\mathbf{\Omega}}_1^{(1)}(t)$ from a construction with an $a_1(t)$ tailored for this component (discarding $\bar{\mathbf{\Omega}}_2^{(1)}(t)$ in this case), $\bar{\mathbf{\Omega}}_2^{(2)}(t)$ from a different $a_2(t)$ (discarding $\bar{\mathbf{\Omega}}_1^{(2)}(t)$), and finally join these two vectors into the final Omn\`es matrix $\Omega(t)=\big\{\bar{\mathbf{\Omega}}_1^{(1)}(t),\bar{\mathbf{\Omega}}_2^{(2)}(t)\big\}$. 

Below the two-kaon threshold we take $a_1(t)=a_2(t)=\mathds{1}$, while for $t\geq \tK$
\begin{align}
a_1(t)&=\begin{pmatrix}
        1 & \frac{N(t)}{2|g(t)|\sigma_t^\pi}f(t)\\
\frac{N(t)}{2|g(t)|\sigma_t^K}f(t) & 1-f(t)+\frac{N^2(t)}{1-\eta^2(t)}f(t)
       \end{pmatrix}\ec\nt\\
a_2(t)&=\begin{pmatrix}
        1 & -\frac{2|g(t)|\sigma_t^K}{N(t)}f(t)\\
-\frac{2|g(t)|\sigma_t^\pi}{N(t)}f(t) & 1-f(t)+\frac{1-\eta^2(t)}{N^2(t)}f(t)
       \end{pmatrix}\ec
\end{align}
with
\beq
N(t)=\sqrt{1-\eta(t)^2\cos^2(2\delta(t)-\psi(t))}-\eta(t)\sin(2\delta(t)-\psi(t))\ec
\eeq
cf.~\eqref{ratio_Omij}, and a function $f(t)$ fulfilling $f(\tK)=0$, $f(\tm)=1$, proves adequate. This construction makes sure that $\bar \Omega_{ij}(t)$ is continuous at $\tm$ and that the relations~\eqref{ratio_Omij} hold. 

As we have seen in the previous section, the Omn\`es function is not unique, and therefore there is a priori no reason why the determinant of the resulting $\Omega(t)$ should match the single-channel expectation: we can always multiply $\bar{\mathbf{\Omega}}_i(t)$ with a real function $g(t)$ with $g(0)=1$ without vitiating the above construction. Since the unitarity condition in its form analogous to~\eqref{unitarity} alone implies that
\beq
\det\Omega=\frac{2e^{i\psi}}{\eta\cos(2\delta-\psi)+\cos\psi}\big(\Re\Omega_{11}\Re\Omega_{22}-\Re\Omega_{12}\Re\Omega_{21}\big)\ec
\eeq
and thus ensures the correct phase, this could be used to make the modulus of the determinant coincide with~\eqref{det_fin}. We simply take
\beq
\label{fOmnes}
f(t)=\bigg(\frac{t-\tK}{\tm-\tK}\bigg)^6\ep
\eeq
This choice of the exponent ensures that $f(t)$ decreases rapidly below $\tm$, but not so fast as to cause numerical problems in the matching-point region. In this way, the single-channel result for the determinant is accurately reproduced except for the energy region close to $\tm$.

\section{Application to pion--nucleon and kaon--nucleon scattering}
\label{sec:pin_appl}

The $\pi\pi\to\bar N N$ and $\bar K K\to\bar N N$ $S$-waves $f^0_+(t)$ and $h^0_+(t)$, respectively, fulfill the unitarity relations (cf.~\cite{RS_DHKM} and Appendix~\ref{app:KN})
\begin{align}
\label{unitarity_KN}
 \Im f^0_+(t)&=\sigma^\pi_t\big(t^0_0(t)\big)^*f^0_+(t)\,\theta\big(t-\tpi\big)
+\frac{2}{\sqrt{3}}\,\sigma^K_t\big(g^0_0(t)\big)^*h_+^0(t)\,\theta\big(t-\tK\big)\ec\nt\\
\Im h^0_+(t)&=\sigma^K_t\big(r^0_0(t)\big)^*h^0_+(t)\,\theta\big(t-\tK\big)
+\frac{\sqrt{3}}{2}\,\sigma^\pi_t\big(g^0_0(t)\big)^*f_+^0(t)\,\theta\big(t-\tpi\big)\ec
\end{align}
where $t^0_0$, $g^0_0$, and $r^0_0$ denote the $\pi\pi$, $\pi\pi\to\bar K K$, and $\bar K K$ $S$-waves with $t$-channel isospin $I_t=0$. Since by virtue of unitarity in the $\pi\pi$/$\bar K K$ system we may identify
\beq
t^0_0(t)=\frac{\eta(t) e^{2i\delta(t)}-1}{2i\sigma_t^\pi}\ec\qquad g^0_0(t)=|g(t)|e^{i\psi(t)}\ec \qquad 
r^0_0(t)=\frac{\eta(t) e^{2i(\psi(t)-\delta(t))}-1}{2i\sigma_t^K}\ec
\eeq
the unitarity relation~\eqref{unitarity_KN} reduces to
\beq
\Im \ff(t)=T^*(t)\Sigma(t)\ff(t)\ec\qquad 
\ff(t)=\begin{pmatrix}
               f^0_+(t)\\
\frac{2}{\sqrt{3}}h^0_+(t)
              \end{pmatrix}\ec
\eeq
and with RS equations providing a dispersion relation of the form
\beq
\label{f_disp_rel}
\ff(t)=\boldsymbol{\Delta}(t)+(\mathbf{a}+\mathbf{b}t)(t-\tN)+\frac{t^2(t-\tN)}{\pi}\int\limits_{\tpi}^\infty\diff t'\frac{\Im\ff(t')}{t'^2(t'-\tN)(t'-t)}\ec
\eeq
where $\tN=4m^2$, the discussion in Sect.~\ref{sec:MOsol_finite_tm} applies. We consider here only the twice-subtracted version of the RS equations for $\pi N$ scattering derived in~\cite{RS_DHKM}. The corresponding inhomogeneity for the $\pi N$ system reads
\begin{align}
\Delta_1(t)&=\tilde\Delta^0_+(t)=\hat N^0_+(t)+\bar\Delta^0_+(t)\ec\nt\\
\bar\Delta^0_+(t)&=\frac{1}{\pi}\int\limits^{\infty}_{W_+}\diff W'\sum\limits^\infty_{l=0}
\Big\{\twosub{\tilde G_{0l}}(t,W')\,\Im f^+_{l+}(W')+\twosub{\tilde G_{0l}}(t,-W')\,\Im f^+_{(l+1)-}(W')\Big\}\ep
\end{align}
Here, $\hat N^0_+(t)$ denotes the nucleon pole term, $f^I_{l\pm}(W)$ the $s$-channel partial waves with total angular momentum $j=|l\pm 1/2|$ and isospin index $I=\pm$, and $W_+=m+\mpi$. Moreover, we have already neglected the contributions from higher $t$-channel partial waves, which were shown to be insignificant in~\cite{RS_DHKM}. In fact, the corresponding equation for $\Delta_2(t)$ is very similar, as long as we restrict ourselves to combinations of isospin and angular momentum that couple to $\pi\pi$ (cf.\ Appendix~\ref{app:KN}): the nucleon pole terms need to be replaced by the hyperon-pole contributions~\eqref{hyperon_pole}, $\mpi$ by $\mK$, and the Clebsch--Gordan coefficients in the relation between the $s$-channel amplitudes in $I=\pm$ and $I_s=0,1$ bases are different from $\pi N$ (where $I_s=1/2,\,3/2$). In particular, the kernel functions $\tilde G_{0l}(t,W)$ require no further modification besides $\mpi\to\mK$.    
Finally, the subtraction constants $a_1$ and $b_1$ can be related to standard $\pi N$ subthreshold parameters
\beq
a_1=-\frac{1}{16\pi}\Bigg(\frac{g^2}{m}+d_{00}^++b_{00}^+\frac{\mpi^2}{3}\Bigg)\ec\qquad 
b_1= -\frac{1}{16\pi}\Bigg(d_{01}^+-\frac{b_{00}^+}{12}\Bigg)
\eeq
(see~\cite{RS_DHKM} for precise definitions), and similarly for $a_2$, $b_2$, and $KN$ subthreshold parameters.

Starting from the dispersion relation~\eqref{f_disp_rel}, the solution for $\ff(t)$ can be derived along the lines that led to~\eqref{omnes_fin}. In addition, we may use the spectral representation of the inverse of the Omn\`es matrix
\beq
 \Omega^{-1}(t)=\unity+\frac{t}{\pi}\int\limits_{\tpi}^{\tm}\diff t'\frac{\Im\Omega^{-1}(t')}{t'(t'-t)}
=\unity-t\,\dot\Omega(0)+\frac{t^2}{\pi}\int\limits_{\tpi}^{\tm}\diff t'\frac{\Im\Omega^{-1}(t')}{t'^2(t'-t)}
\eeq
to perform the integrals involving $\mathbf{a}$ and $\mathbf{b}$ explicitly ($\dot\Omega$ denotes the derivative with respect to $t$). In this way, we arrive at
\begin{align}
\label{final_omnes_sol}
\ff(t)&=\boldsymbol{\Delta}(t)+(t-\tN)\Omega(t)(\unity-t\,\dot\Omega(0))\mathbf{a}+t(t-\tN)\Omega(t)\mathbf{b}\nt\\
&\qquad-\frac{t^2(t-\tN)}{\pi}\Omega(t)\int\limits_{\tpi}^{\tm}\diff t'\frac{\Im\Omega^{-1}(t')\boldsymbol{\Delta}(t')}{t'^2(t'-\tN)(t'-t)}
+\frac{t^2(t-\tN)}{\pi}\Omega(t)\int\limits_{\tm}^\infty\diff t'\frac{\Omega^{-1}(t')\Im \ff(t')}{t'^2(t'-\tN)(t'-t)}\ep
\end{align}

To obtain the corresponding numerical results for $f^0_+(t)$ and $h^0_+(t)$ we now collect the various input needed for the explicit evaluation of~\eqref{final_omnes_sol}. We take the $\pi\pi$ phase and inelasticity from the extended Roy-equation analysis of~\cite{CCL:Regge,CCL:PWA} and the $\pi\pi\to\bar K K$ partial wave from~\cite{BDM04}, where, in the pseudophysical region $\tpi\leq t\leq \tK$, the modulus $|g(t)|$ was determined as the solution of RS equations for $\pi K$ scattering, while above the two-kaon threshold phase-shift solutions~\cite{Cohen,Etkin} were used.\footnote{ We are indebted to Bachir Moussallam for providing a version of the solution for $g(t)$ consistent with the $\pi\pi$ phase shift of~\cite{CCL:Regge,CCL:PWA}.} The two-channel approximation in terms of $\pi\pi$ and $\bar K K$ intermediate states should work well at lower energies where the $f_0(980)$ resonance dominates, but will break down once inelasticities arising from $4\pi$ intermediate states become important, which are expected to set in around $\sqrt{t_0}=1.3\,{\rm GeV}$. Evidently, this still leaves a large part of the $\pi\pi\to\bar N N$ pseudophysical region $\tpi\leq t\leq \tN$  where intermediate states besides those considered here may contribute significantly to the spectral function. However, two subtractions should already appreciably suppress the sensitivity to this high-energy regime, which we will examine in more detail below.

In view of the remaining uncertainties in the pseudophysical region we refrain from including information on $f^0_+(t)$ even above $\tN$ from a phase-shift solution~\cite{Anisovich}, but just put $\Im \ff=0$ above $\tm$. Due to continuity of the MO solution at $\tm$ this implies that the solutions for $f^0_+(t)$ and $h^0_+(t)$ will vanish at $\tm$ as well (unless the phases happen to pass an integer multiple of $\pi$ at $\tm$). We take advantage of the fact that for kinematic reasons $f^0_+(t)$ and $h^0_+(t)$ have a zero at the physical threshold $\tN$ and choose the matching point accordingly as $\tm=\tN$, which should allow for a reasonably smooth matching.

Next, to evaluate the inhomogeneities $\Delta_i(t)$ we need the imaginary parts for the $s$-channel partial waves of $\pi N$ and $KN$ scattering. As discussed in~\cite{RS_DHKM}, we rely on the GWU solution~\cite{GWU:2008,SAID} for $W\leq W_{\rm a}=2.5\,{\rm GeV}$ in the $\pi N$ case, summing up all partial waves up to $l_{\rm max}=4$, and on the Regge model~\cite{piNuRegge} for $W>W_{\rm a}$. Similarly, $\Delta_2$ is evaluated based on~\cite{SAID,GWU:KN}, integrating the partial waves with $l\leq4$ up to $2\,{\rm GeV}$. 

\begin{figure}[t!]
\centering
\includegraphics[width=0.49\textwidth]{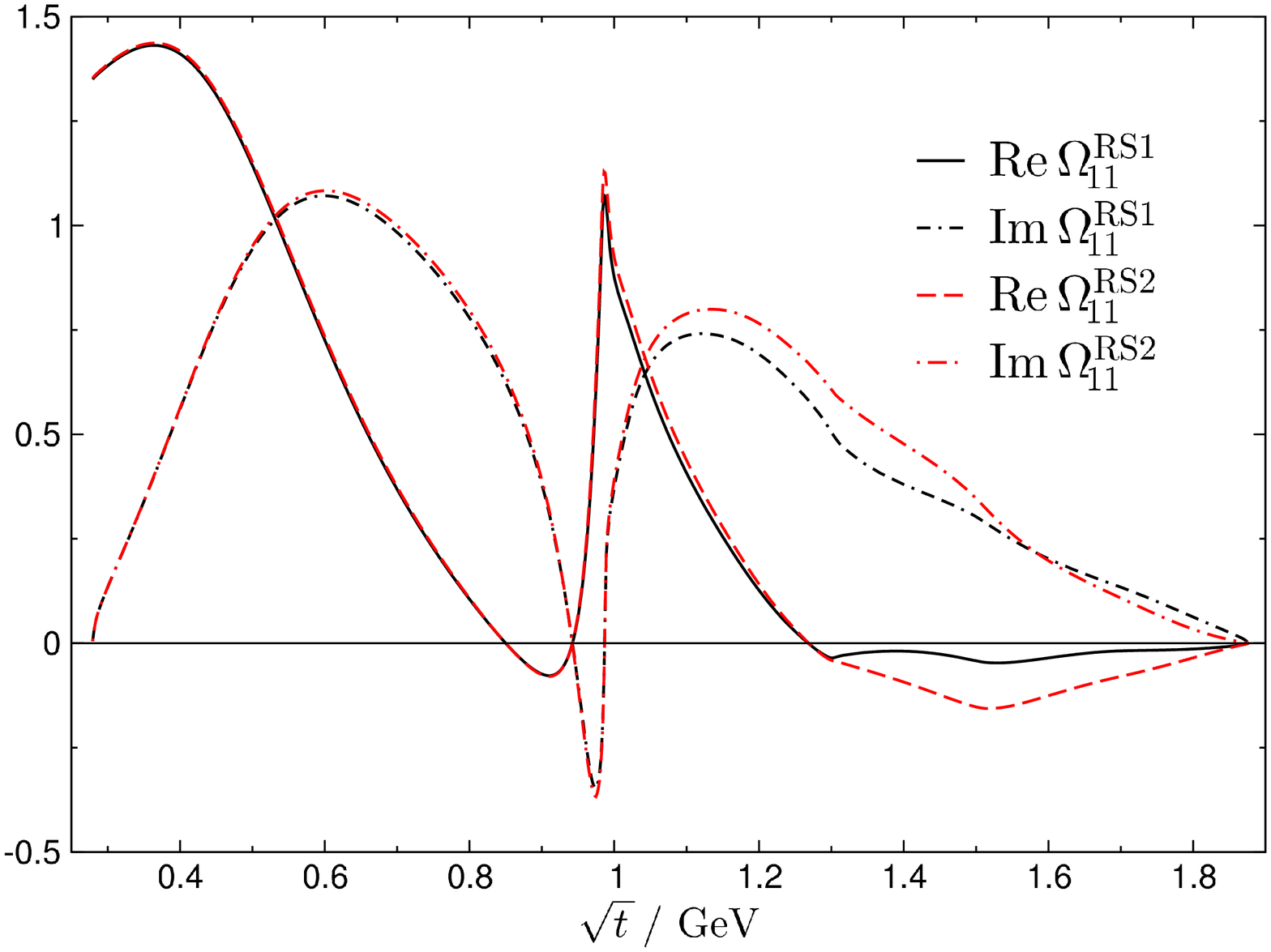}
\includegraphics[width=0.49\textwidth]{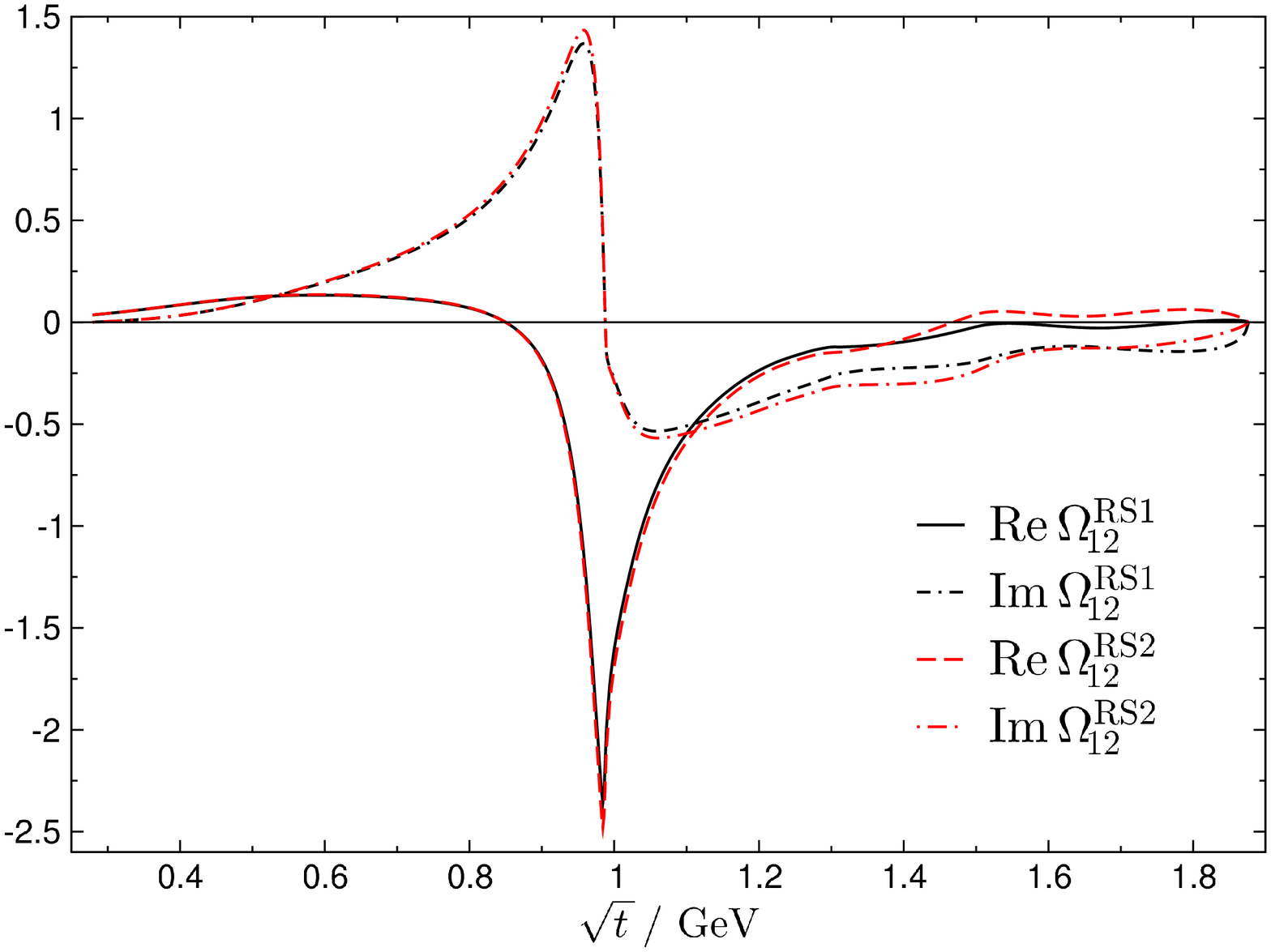}\\
\includegraphics[width=0.49\textwidth]{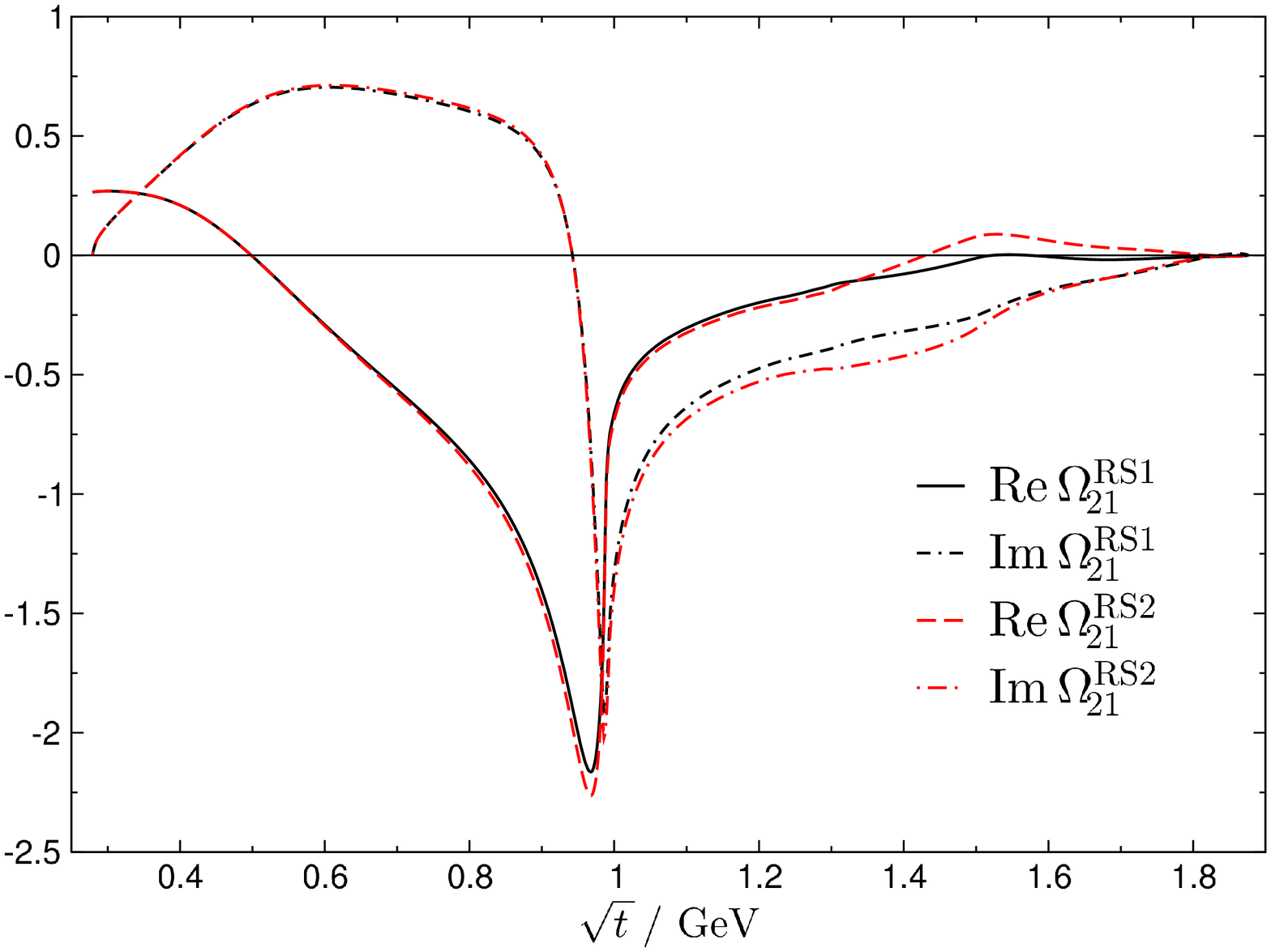}
\includegraphics[width=0.49\textwidth]{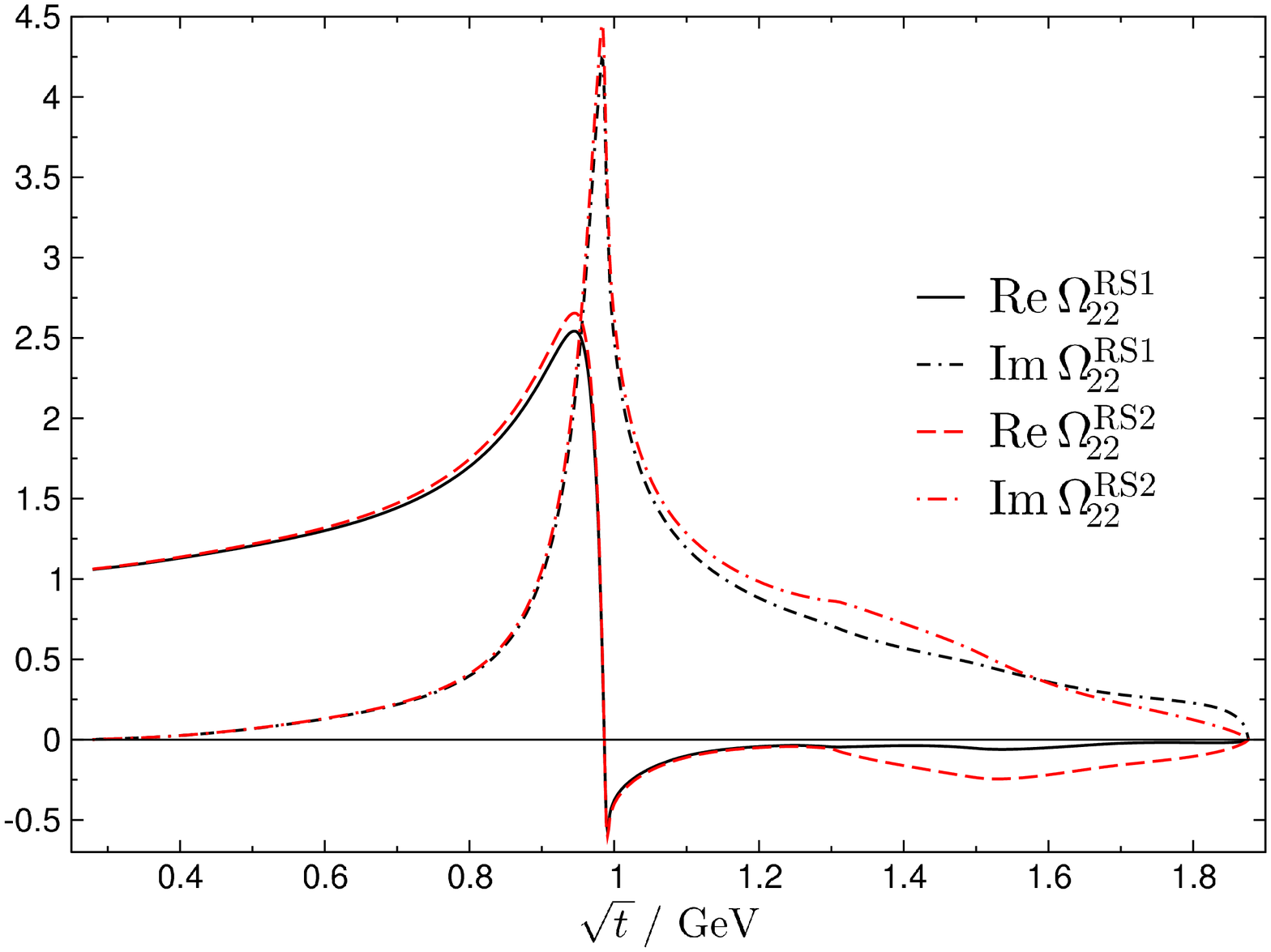}
\caption{Results for real and imaginary part of the components of the Omn\`es matrix for the input phases RS1 and RS2 as described in the main text.}
\label{fig:Omnesmatrix}
\end{figure} 

\begin{figure}
\centering
\includegraphics[width=0.49\textwidth]{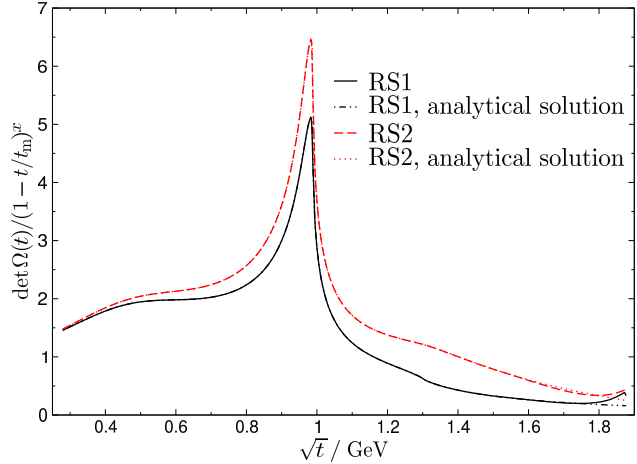}
\caption{Comparison between numerical and analytical result for the determinant of the Omn\`es matrix.}
\label{fig:Omnesdet}
\end{figure} 

In the remainder of the paper we will consider three variants of the input described above. First, we take $\delta$ and $\psi$ to be constant above $t_0$ (``RS1''), second, we guide $\delta$ and $\psi$ smoothly to $2\pi$ above $t_0$ by means of~\cite{Moussallam99}
\beq
\label{phase_extrapolation}
\delta(t)=2\pi+(\delta(t_0)-2\pi)\hat f\Big(\frac{t}{t_0}\Big)\ec\qquad \hat f(x)=\frac{2}{1+x^{3/2}}\ec 
\eeq
keeping the phase constant above $\tm$ (``RS2''), and third, we modify RS1 in such a way that $\Delta_2=0$ (``RS3''). In all three cases, $|g(t)|$ is led smoothly to zero above $\tm$ by a prescription similar to~\eqref{phase_extrapolation}. The choice of these variants is motivated as follows. As indicated above, the model for the $\pi\pi$/$\bar K K$ $S$-matrix is only meaningful roughly up to $t_0$, and ideally our results should be insensitive to variations of this input above $t_0$, the simplest choice of course being to keep the phases constant. However, to ensure the correct asymptotic behavior for the scalar meson form factors, the phase $\psi$ must tend to an asymptotic value of $2\pi$~\cite{Moussallam99}, which, phenomenologically, also suggests to guide $\delta$ to $2\pi$. Thus, RS1 and RS2 are convenient choices to assess the sensitivity to the high-energy input for the phase shifts. The results for the Omn\`es matrix corresponding to these two scenarios are depicted in Fig.~\ref{fig:Omnesmatrix}. In addition, we compare the results for $\det\Omega(t) /(1-t/\tm)^x$ to the analytically known results in Fig.~\ref{fig:Omnesdet}. As expected, the only deviations occur in the proximity of $\tm$, where the modifications originating from $f(t)$ according to~\eqref{fOmnes} set in. Note that the difference between RS1 and RS2 appears slightly exaggerated here, since $x$ differing in both cases leads to a different factor being divided out.  

 Finally, we use the reference point~\cite{Hoehler}
\beq
\label{subthresh_coupling}
\frac{g^2}{4\pi}=14.28\ec\qquad d_{00}^+=-1.46\,\mpi^{-1}\ec\qquad d_{01}^+=1.14\,\mpi^{-3}\ec\qquad b_{00}^+=-3.54\,\mpi^{-3}\ec
\eeq
although we will display the dependence of $\Delta_\sigma$ on each of these parameters explicitly in the end. In contrast, we are not aware of reliable input for the $KN$ subthreshold parameters, and simply put $a_2=b_2=0$. In fact, this approximation gives reason to investigate RS3, since the results from the $\pi N$ sector show that the contributions from the corresponding parameters will certainly not be larger than the sum of $KN$ Born terms (conventions for masses and couplings of the hyperons are given in Appendix~\ref{app:KN}) and $s$-channel integrals. The difference between RS1 and RS3 thus serves as an estimate of the uncertainty induced by neglecting the $KN$ subthreshold parameters.

\begin{figure}[t!]
\centering
\includegraphics[width=0.49\textwidth]{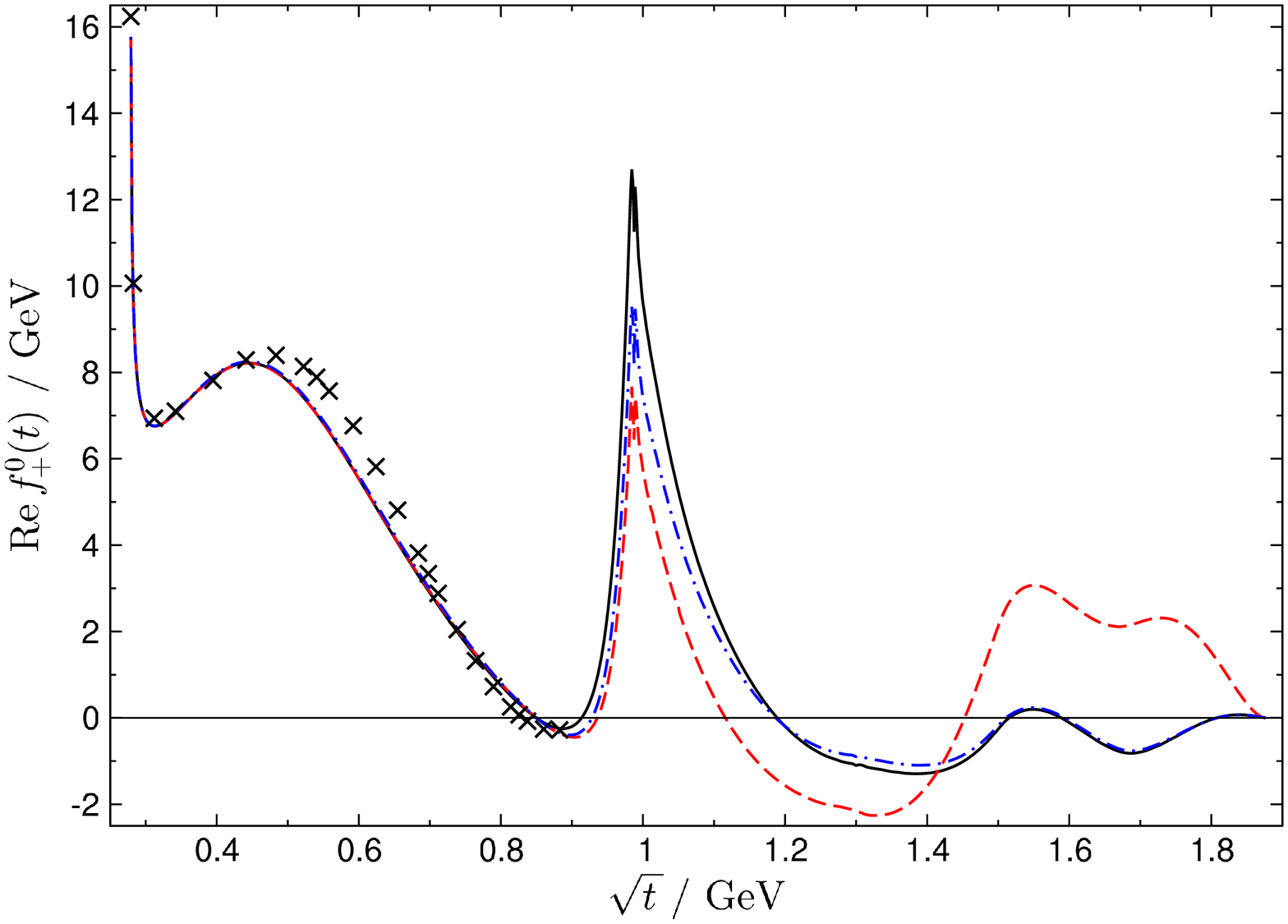}
\includegraphics[width=0.49\textwidth]{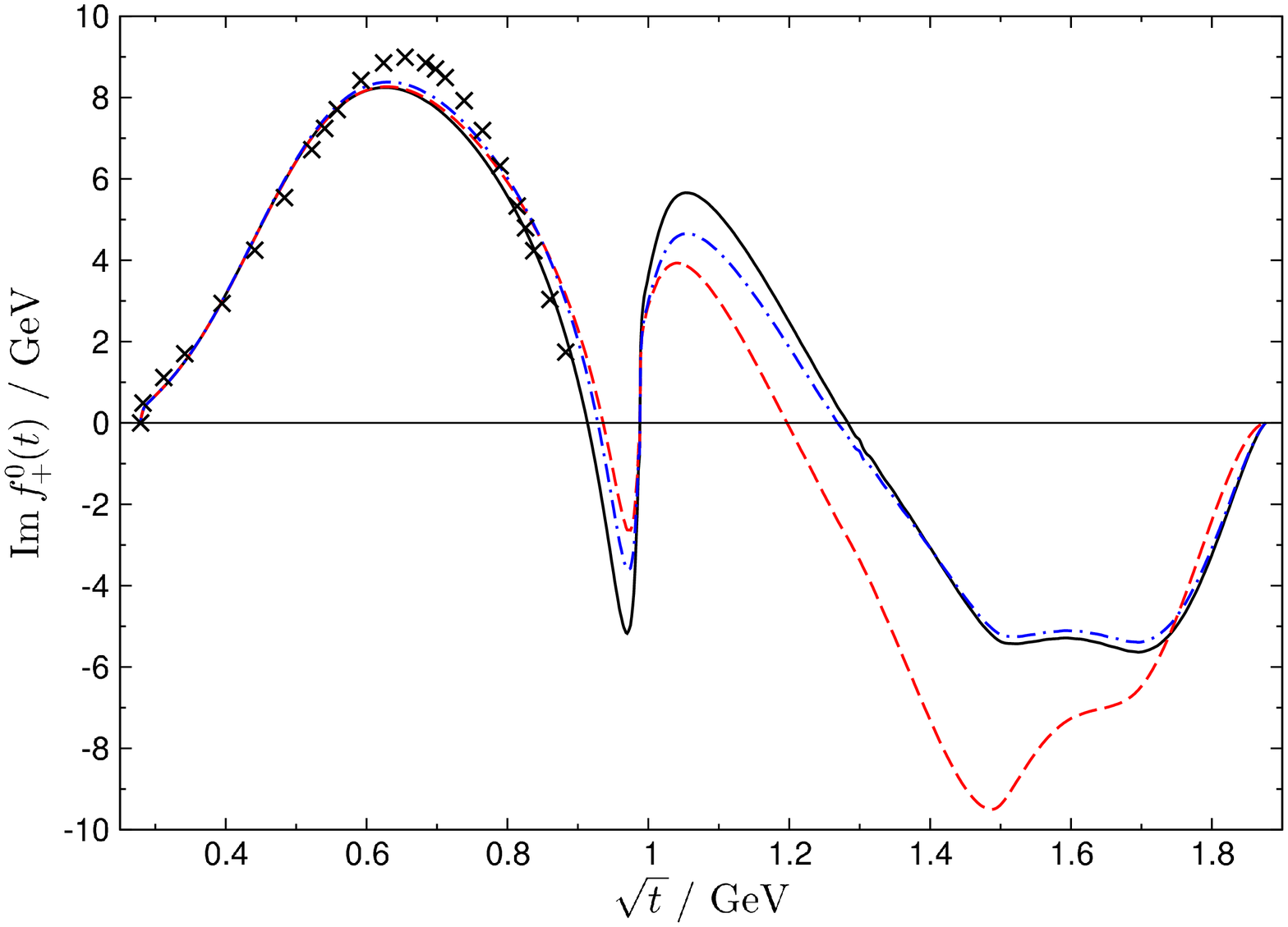}\\
\includegraphics[width=0.49\textwidth]{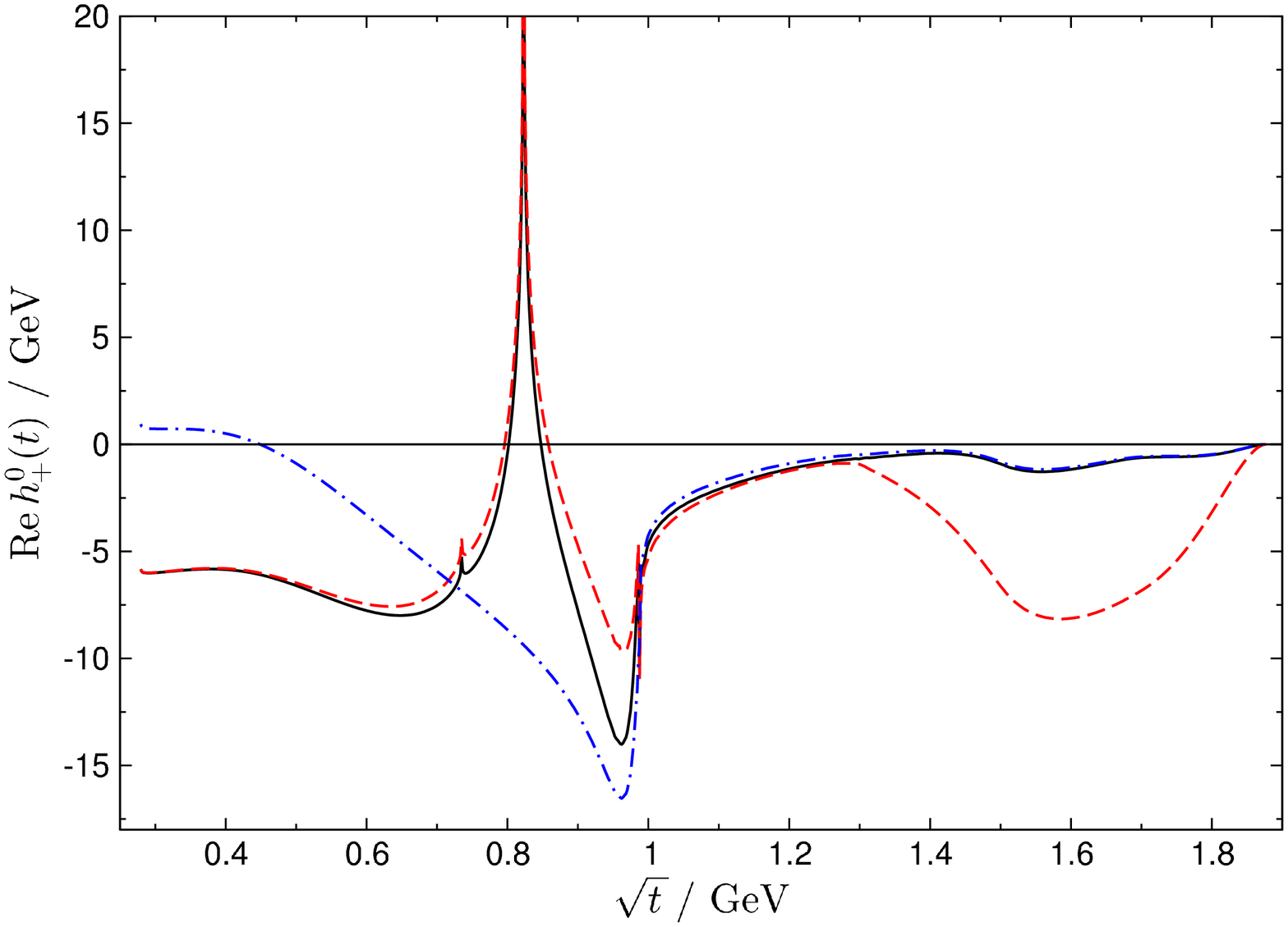}
\includegraphics[width=0.49\textwidth]{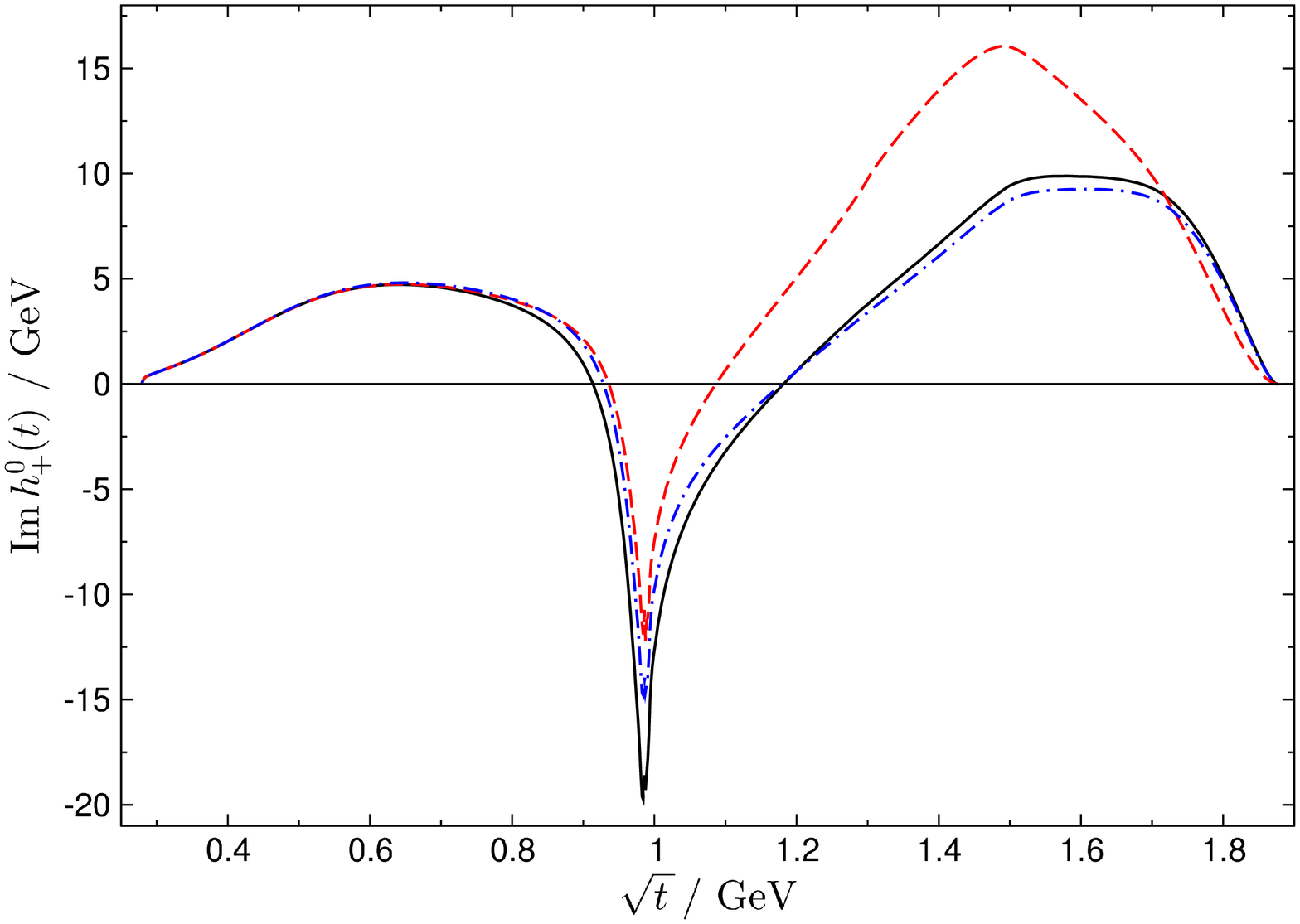}
\caption{Results for real and imaginary part of $f^0_+(t)$ and $h^0_+(t)$. The solid, dashed, and dot-dashed lines refer to the input RS1, RS2, and RS3 as described in the main text. For $f^0_+(t)$ the black crosses indicate the results of~\cite{Hoehler}.}
\label{fig:f0ph0p}
\end{figure} 

The results for $f^0_+(t)$ and $h^0_+(t)$ are shown in Fig.~\ref{fig:f0ph0p}. In the case of $f^0_+(t)$ the agreement between the different parameter sets is very good up to $1\,{\rm GeV}$. It is striking that the difference between RS1 and RS2 is much larger than between RS1 and RS3, which indicates that the results are much more sensitive to the choice of the phases beyond $t_0$ than to the details of the $KN$ amplitude. The real part of $h^0_+(t)$ exhibits two distinct divergences below $1\,{\rm GeV}$ that correspond to the pole-term contributions from the $\Lambda$ (large peak at $0.82\,{\rm GeV}$) and $\Sigma$ (small peak at $0.73\,{\rm GeV}$) hyperon, strictly analogous to the nucleon pole in $f^0_+(t)$ that emerges slightly below $\tpi$. These poles, which disappear in RS3 by construction, do not pose a problem in practice, since $h^0_+(t)$ only contributes to the spectral function of the scalar form factor of the nucleon above $\tK$. Apart from these poles below the two-kaon threshold the conclusion is very similar to $f^0_+(t)$: the uncertainty in the phase shifts outweighs the uncertainty in the $KN$ input.

\section{Scalar form factors}
\label{sec:SFF}

\subsection{Scalar pion and kaon form factors}

We define the scalar pion and kaon form factors as
\beq
F^S_\pi(t)=\langle \pi(p')|\hat m(\bar u u+\bar d d)|\pi(p)\rangle\ec\qquad F^S_K(t)=\langle K(p')|\hat m(\bar u u+\bar d d)|K(p)\rangle\ec\qquad t=(p'-p)^2\ep
\eeq
In the two-channel approximation they fulfill the unitarity relation~\cite{DGL90}
\beq
\Im\mathbf{F}^S(t)=T^*(t)\Sigma(t)\mathbf{F}^S(t)\ec\qquad \mathbf{F}^S(t)=\begin{pmatrix}
                                                                                  F^S_\pi(t)\\ \frac{2}{\sqrt{3}}F^S_K(t)
                                                                                 \end{pmatrix}\ec
\eeq
and thus
\beq
\mathbf{F}^S(t)=\alpha \mathbf{\Omega}_1^\infty+\beta \mathbf{\Omega}_2^\infty\ec
\eeq
with the infinite-matching-point Omn\`es solutions $\mathbf{\Omega}_i^\infty$  defined in~\eqref{Omnes_vector}. The phases $\delta$ and $\psi$ are guided smoothly to their assumed asymptotic value of $2\pi$ according to~\eqref{phase_extrapolation}. Using the normalization $\Omega^\infty(0)=\unity$ of the Omn\`es matrix to pin down the coefficients $\alpha$, $\beta$, we find
\begin{align}
\label{SFF_Omnes}
 F^S_\pi(t)&=F^S_\pi(0)\Omega_{11}^\infty(t)+\frac{2}{\sqrt{3}}F^S_K(0)\Omega_{12}^\infty(t)\ec\nt\\
F^S_K(t)&=\frac{\sqrt{3}}{2}F^S_\pi(0)\Omega_{21}^\infty(t)+F^S_K(0)\Omega_{22}^\infty(t)\ep
\end{align}
The form factors at $t=0$ can be determined via the Feynman--Hellmann theorem~\cite{Hellmann,Feynman}
\beq
F^S_\pi(0)=\hat m\frac{\partial}{\partial \hat m}\mpi^2\ec\qquad
F^S_K(0)=\hat m\frac{\partial}{\partial \hat m}\mK^2\ec
\eeq 
from the quark-mass dependence of the meson masses. For the pion form factor the result at $\Order(p^4)$ in the chiral expansion reads~\cite{GL84}
\beq
F^S_\pi(0)=\mpi^2-\frac{\mpi^4}{32\pi^2\Fpi^2}(\bar l_3-1)=(0.984\pm0.006)\mpi^2\ec
\eeq 
where we have used $\bar l_3=3.2\pm 0.8$ \cite{FLAG}. The leading-order result for the kaon form factor
\beq
F^S_K(0)=\frac{\mpi^2}{2}
\eeq
is subject to potentially large $SU(3)$ corrections, which in the isospin limit amount to~\cite{GL85}
\begin{align}
 F^S_K(0)&=\frac{\mpi^2}{2}\Bigg\{1+\frac{\meta^2}{32\pi^2\Fpi^2}\log\frac{\meta^2}{\mu^2}+\frac{\mK^2}{72\pi^2\Fpi^2}\bigg(\log\frac{\meta^2}{\mu^2}+1\bigg)-\frac{\mpi^2}{32\pi^2\Fpi^2}\log\frac{\mpi^2}{\mu^2}\nt\\
&\qquad+\frac{8(2\mK^2-\mpi^2)}{\Fpi^2}\big(2L_8^{\rm r}-L_5^{\rm r}\big)+\frac{32\mK^2}{\Fpi^2}\big(2L_6^{\rm r}-L_4^{\rm r}\big)\Bigg\}\ec
\end{align}
where $\meta$ is the mass of the $\eta$, $\Fpi$ the pion decay constant, and $\mu$ the renormalization scale.
Varying the low-energy constants in the range~\cite{FLAG}
\beq
2L_8^{\rm r}-L_5^{\rm r}=(-0.35\ldots +0.1)\cdot 10^{-3}\ec\qquad 2L_6^{\rm r}-L_4^{\rm r}=(0\ldots +0.2)\cdot 10^{-3}\ec
\eeq
corresponds to $F^S_K(0)=(0.4\ldots0.6)\,\mpi^2$. In the following, we will restrict $F^S_K(0)$ to lie within these boundaries, while adopting $F^S_K(0)=\mpi^2/2$ as our central solution.

\begin{figure}[t!]
\centering
\includegraphics[height=0.268\textheight]{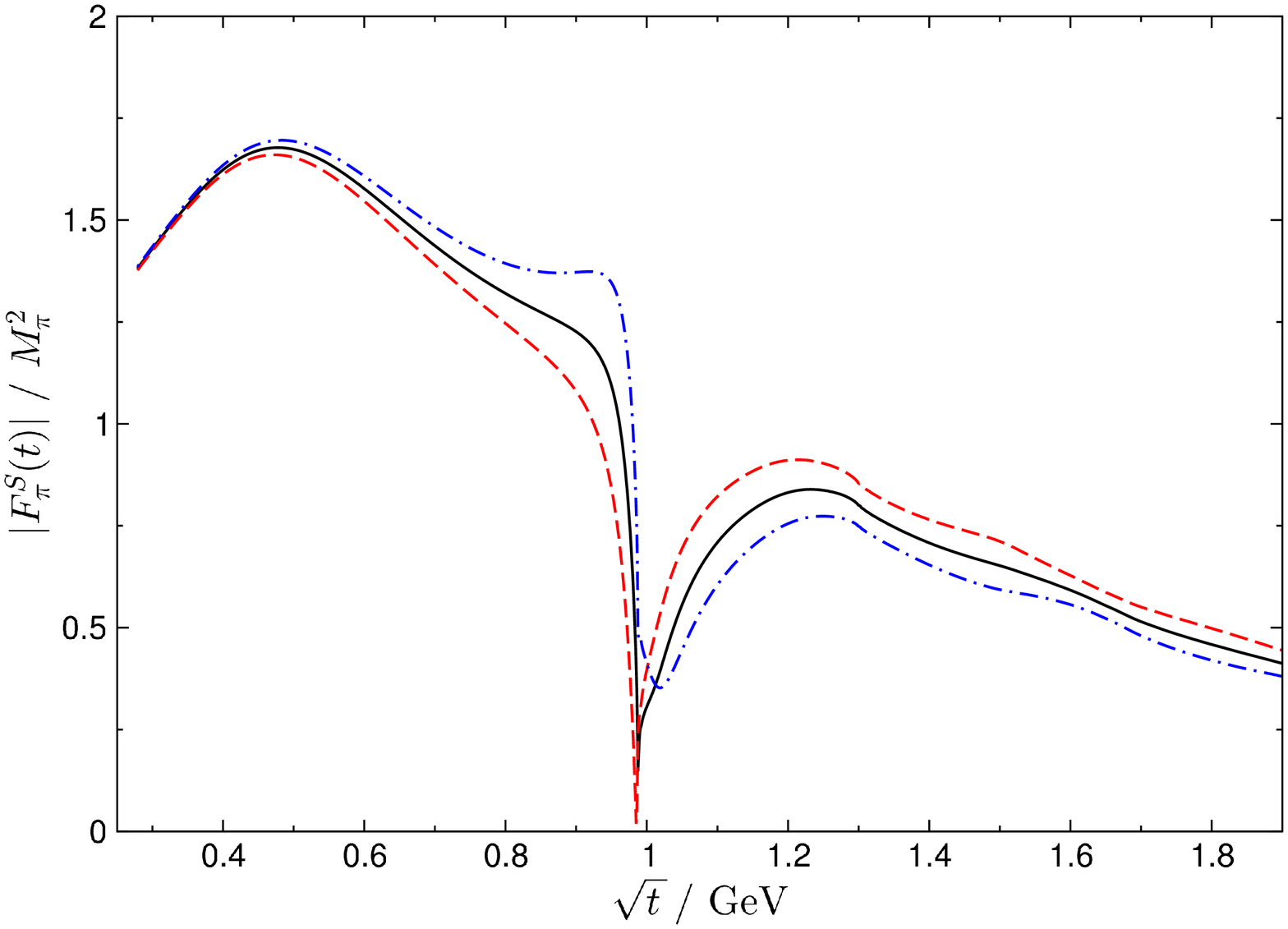}
\includegraphics[height=0.268\textheight]{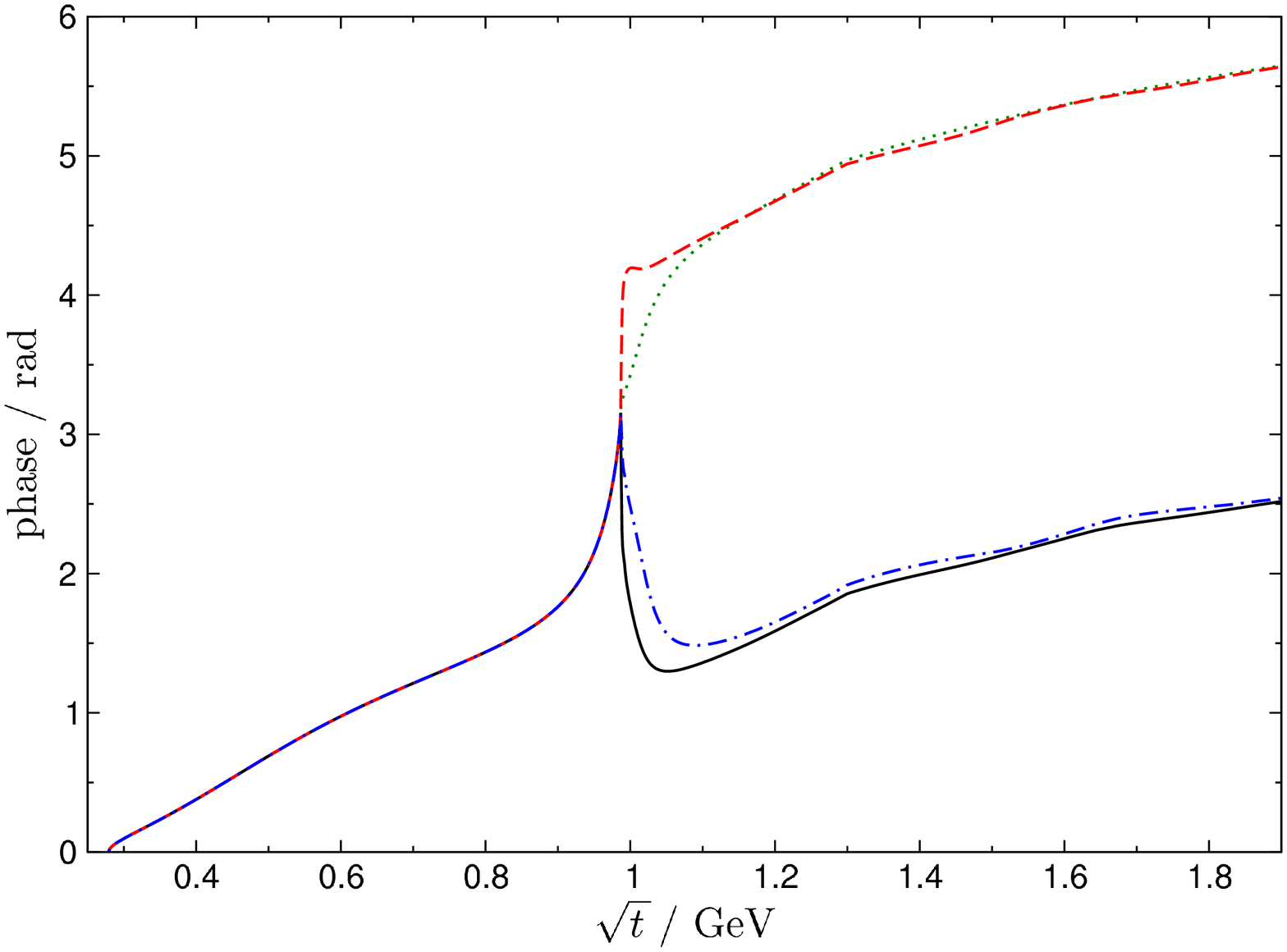}\\
\includegraphics[height=0.268\textheight]{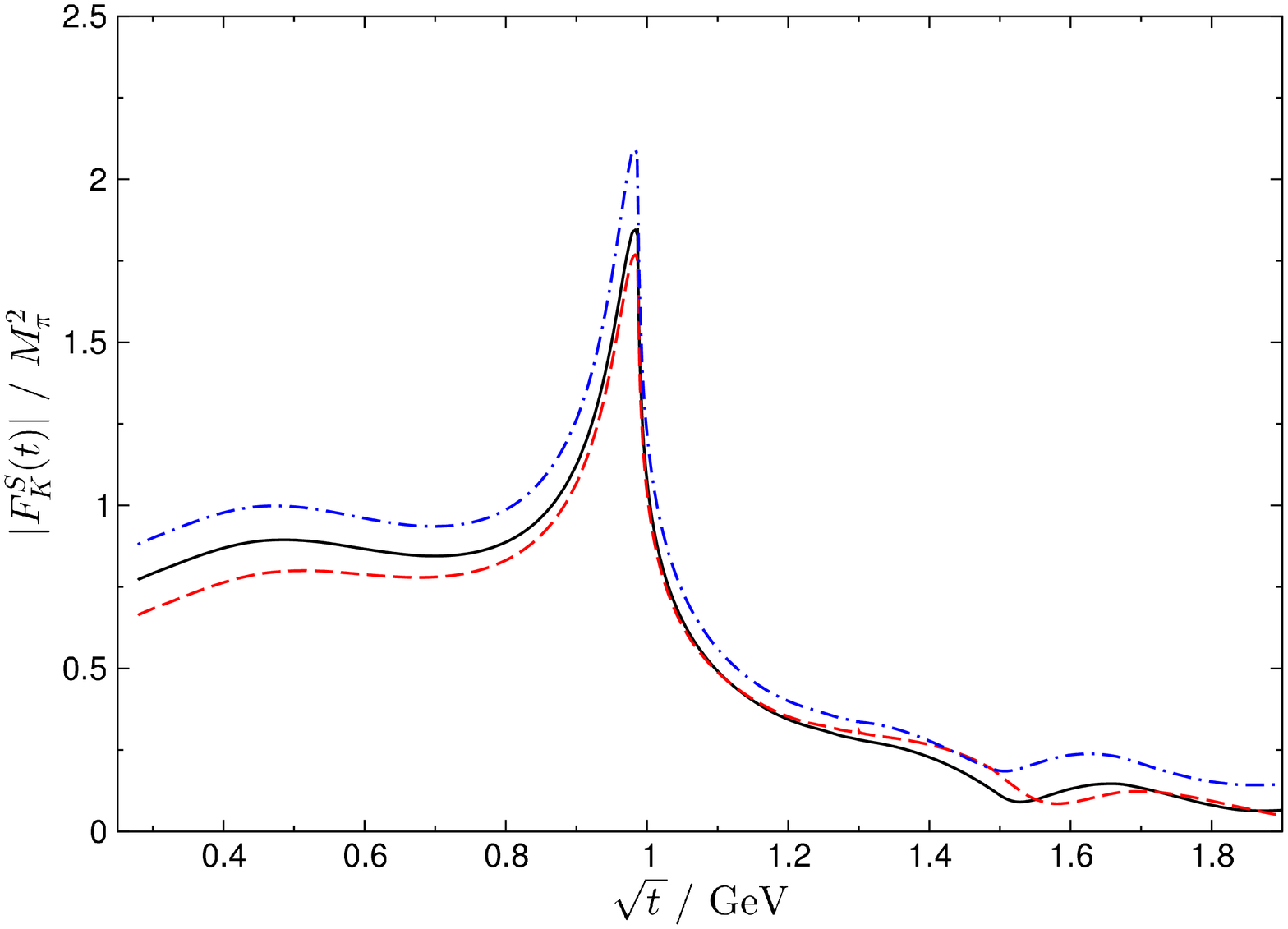}
\includegraphics[height=0.265\textheight]{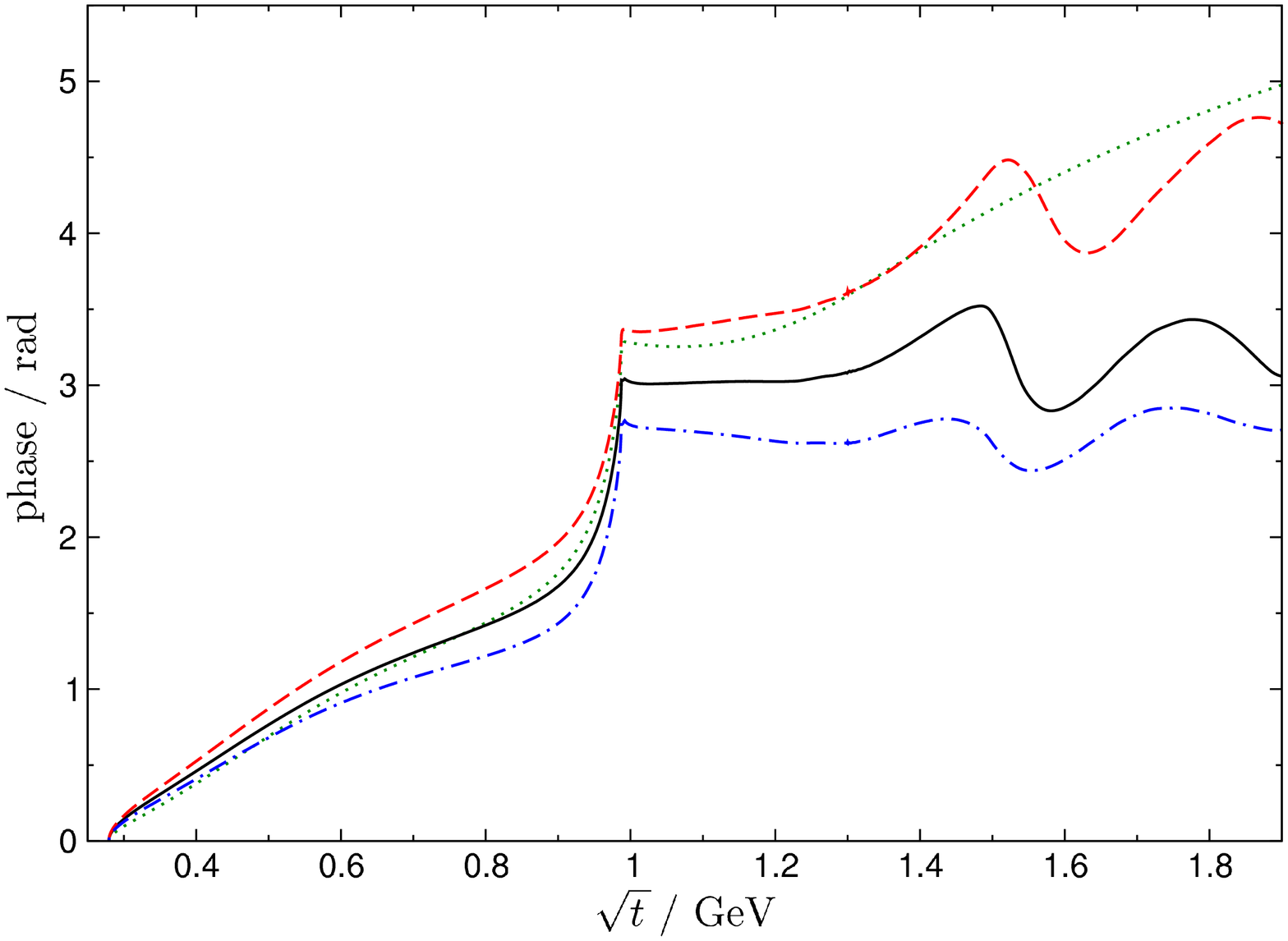}
\caption{Modulus (left) and phase (right) of the scalar pion and kaon form factors. The solid, dashed, and dot-dashed lines refer to $F^S_K(0)=\mpi^2/2$, $0.4\,\mpi^2$, and $0.6\,\mpi^2$. The phases of $F^S_\pi$ and $F^S_K$ are compared to $\delta$ and $\psi$, respectively, as indicated by the dotted lines.}
\label{fig:SFF}
\end{figure} 

The corresponding results for the form factors are depicted in Fig.~\ref{fig:SFF}. The phase of $F^S_\pi$ coincides with $\delta$ below $\tK$, as required by unitarity, cf.~\eqref{Omnes_below_tK} and~\eqref{SFF_Omnes}. Above $\tK$, the behavior of the phase actually depends on the assumption for $F^S_K(0)$, it either largely follows $\delta$ or abruptly drops by $\pi$. The appearance of the first scenario is surprising in view of the results of~\cite{ACCGL} (see also~\cite{Oller2007}), where only the second behavior is mentioned (the assumption for $F^S_K(0)$ agrees with our central solution). The reason for this discrepancy can be understood as follows. The $S$-matrix used in~\cite{ACCGL} involves a phase $\delta$ that fulfills $\delta(\tK)<\pi$. In this case, the phase $\delta_t$ of the full $\pi\pi$ partial wave $t^0_0$ itself displays the characteristic drop above $\tK$, reflecting the fact that the phase arrives at the two-kaon threshold immediately before completing a full circle in the Argand diagram. However, in recent years, it seems to have become consensus that $\delta(\tK)>\pi$ is more likely~\cite{CCL:Regge,CCL:PWA,GKPY,Moussallam11}, which implies that $\delta_t$ by no means exhibits a sharp drop above $\tK$. We conclude that the behavior of the phase cannot simply be deduced from the phase of $t^0_0$, it crucially depends on the relative strength $F^S_\pi(0)/F^S_K(0)$ in the superposition of the two terms involving different components of the Omn\`es matrix as given in~\eqref{SFF_Omnes}, thus attesting to the inherent two-channel nature of the problem.\footnote{One immediate consequence is that an effective single-channel Omn\`es description of $F^S_\pi$ in terms of the phase of $F^S_\pi$ will only be applicable for certain ranges in $F^S_\pi(0)/F^S_K(0)$, unless the phase is supplemented by hand with an additional term $-\pi\theta(t-\tK)$.}  From this point of view, it is not surprising that the phase may behave differently if $F^S_K(0)$ is varied.  In contrast, the phase of $F^S_K(t)$ roughly follows the shape of $\psi$ for all three solutions. Note that in this case~\eqref{Omnes_below_tK} does not impose any additional constraints on the phase below $\tK$.

\begin{table}
\renewcommand{\arraystretch}{1.3}
\centering
\begin{tabular}{c r r r}
\toprule
$F^S_K(0)$ & $0.4\,\mpi^2$ & $\mpi^2/2$ & $0.6\,\mpi^2$ \\
\midrule
$\langle r^2\rangle^S_\pi~[{\rm fm}^2]$ & $0.575$ & $0.584$ & $0.592$\\
$\langle r^2\rangle^S_K~[{\rm fm}^2]$ & $0.835$ & $0.710$ & $0.626$\\
\bottomrule
\end{tabular}
\renewcommand{\arraystretch}{1.0}
\caption{Dependence of the scalar pion and kaon radii on $F^S_K(0)$.}
\label{tab:radii}
\end{table}

Finally, we can express the scalar radii in terms of the form factors at $t=0$ and the derivative of the Omn\`es matrix
\begin{align}
\langle r^2\rangle^S_\pi&=6\bigg\{\dot\Omega_{11}^\infty(0)+\frac{2}{\sqrt{3}}\frac{F^S_K(0)}{F^S_\pi(0)}\dot\Omega_{12}^\infty(0)\bigg\}\ec\nt\\
\langle r^2\rangle^S_K&=6\bigg\{\frac{\sqrt{3}}{2}\frac{F^S_\pi(0)}{F^S_K(0)}\dot\Omega_{21}^\infty(0)+\dot\Omega_{22}^\infty(0)\bigg\}\ep
\end{align}
In this way, the derivative of the Omn\`es matrix, e.g.\ for our central solution
\beq
\dot\Omega^\infty(0)=\begin{pmatrix}
                      2.31 & 0.32\\ 1.26 & 0.89
                     \end{pmatrix}
\,{\rm GeV}^{-2}\ec
\eeq
leads to the results for the scalar radii summarized in Table~\ref{tab:radii}. Our results for the scalar pion radius are in good agreement with $\langle r^2\rangle^S_\pi=(0.61\pm0.04)\,{\rm fm^2}$ from~\cite{CGL01} and the range $\langle r^2\rangle^S_\pi=(0.583\ldots 0.653)\,{\rm fm^2}$ found in~\cite{Moussallam99}. Albeit attached with a fairly large uncertainty, the values for $\langle r^2\rangle^S_K$ lie systematically higher than its ChPT expectation $\langle r^2\rangle^S_K\sim 0.3\,{\rm fm}^2$~\cite{FKM02} (for a detailed comparison of the dispersive and the ChPT result as well as the role of $\Order(p^6)$ corrections see~\cite{Bijnens_Dhonte}). In both approaches the uncertainties are substantial, either due to the large sensitivity to the specific input in the dispersive calculation or due to insufficient knowledge of low-energy constants and higher-order corrections. As the precise value of the scalar kaon radius is irrelevant for the present study, we do not consider this issue any further.

\subsection{Scalar form factor of the nucleon}

The scalar form factor of the nucleon is defined as
\beq
\sigma(t)=\frac{1}{2m}\langle N(p')|\hat m(\bar u u+\bar d d)|N(p)\rangle\ec\qquad t=(p'-p)^2\ep
\eeq
It fulfills the once-subtracted dispersion relation
\beq
\sigma(t)=\sigma_{\pi N}+\frac{t}{\pi}\int\limits_{\tpi}^\infty\diff t'\frac{\Im\sigma(t')}{t'(t'-t)}\ec
\eeq
where the pion--nucleon $\sigma$ term $\sigma_{\pi N}=\sigma(0)$ acts as subtraction constant. In this way, evaluation at $t=2\mpi^2$ gives access to $\Delta_\sigma$ (as defined in~\eqref{Delta_sigma_def}), provided that the imaginary part is sufficiently well constrained to perform the dispersive integral. Generalizing the result quoted in~\cite{GLS90_FF} by including $\bar K K$ intermediate states, the spectral function becomes
\beq
\label{unitarity_sigma_term}
\Im\sigma(t)=-\frac{1}{p_t^2\sqrt{t}}\bigg\{\frac{3}{4}q_t\big(F^S_\pi(t)\big)^*f^0_+(t)\,\theta\big(t-\tpi\big)+k_t\big(F^S_K(t)\big)^*h^0_+(t)\,\theta\big(t-\tK\big)\bigg\}\ep
\eeq

\begin{figure}[t!]
\centering
\includegraphics[width=0.49\textwidth]{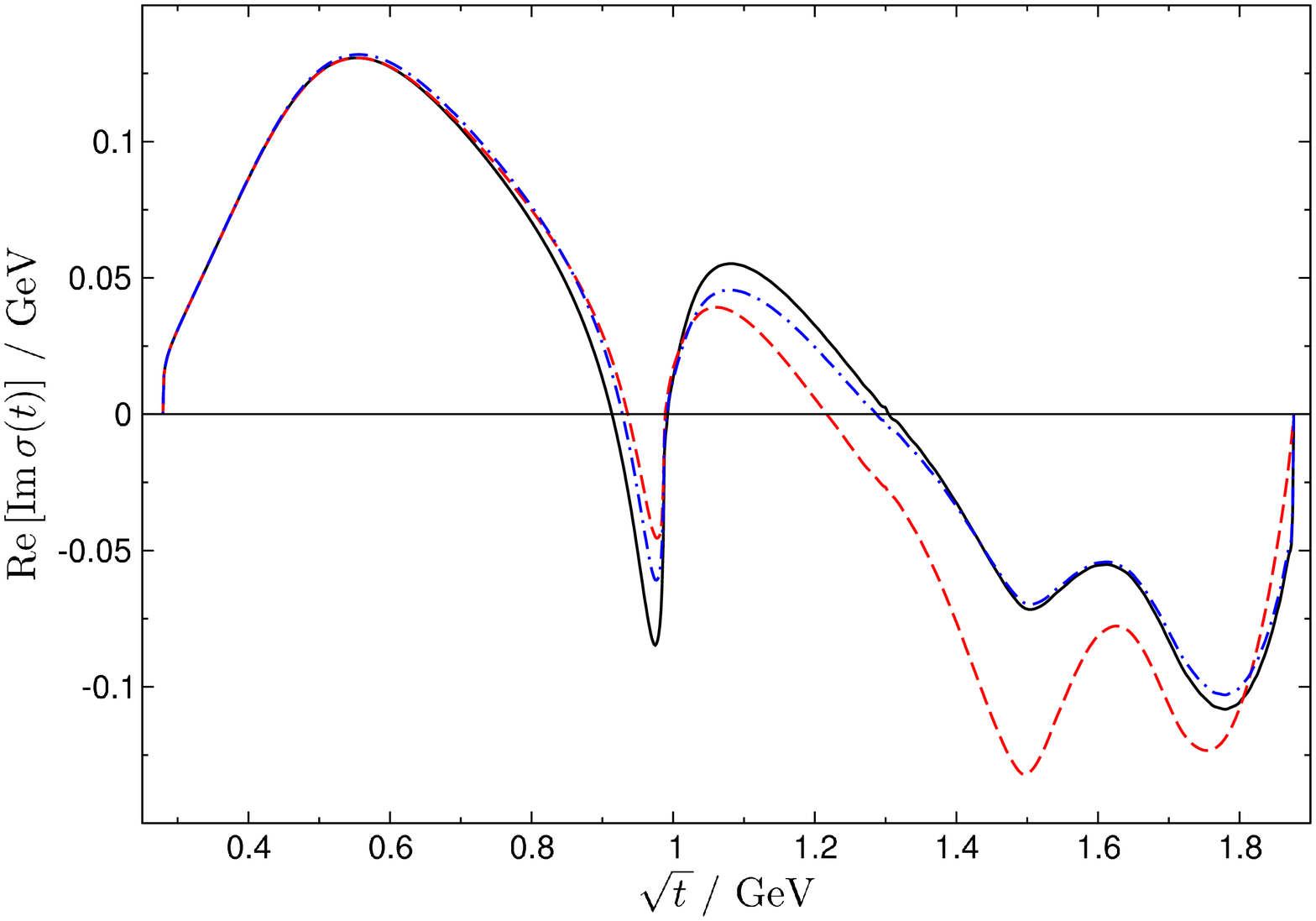}
\includegraphics[width=0.49\textwidth]{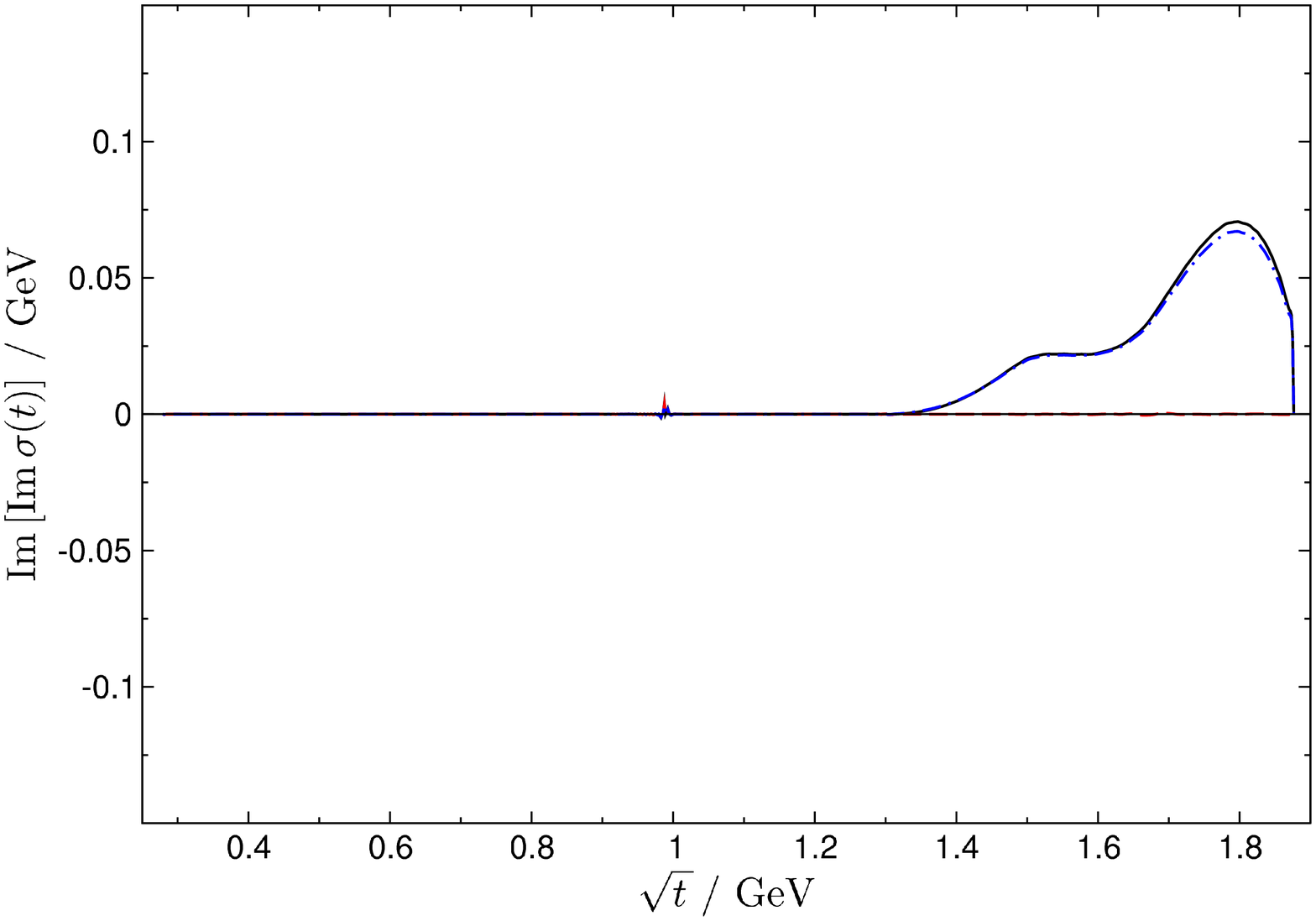}\\
\includegraphics[width=0.49\textwidth]{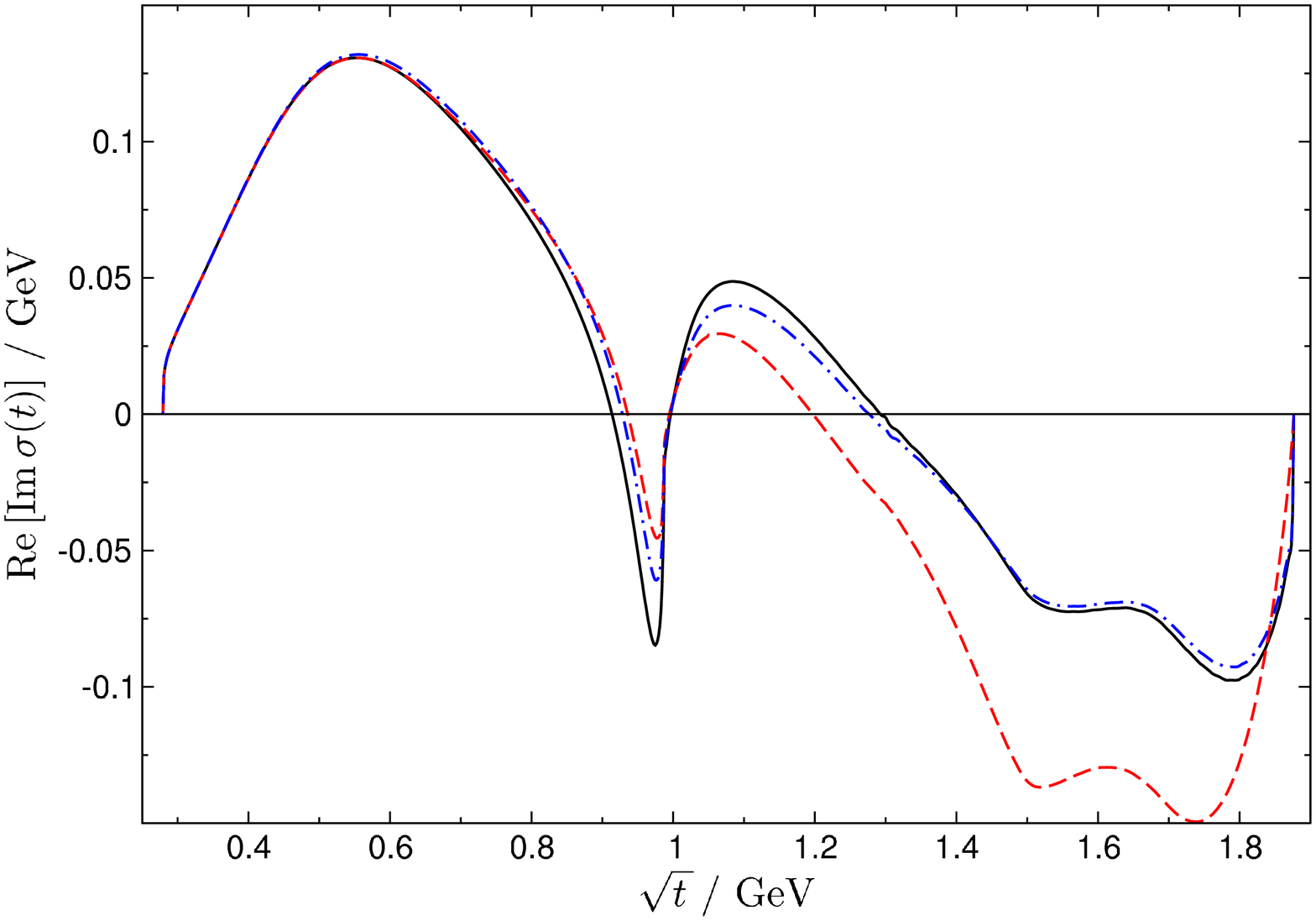}
\includegraphics[width=0.49\textwidth]{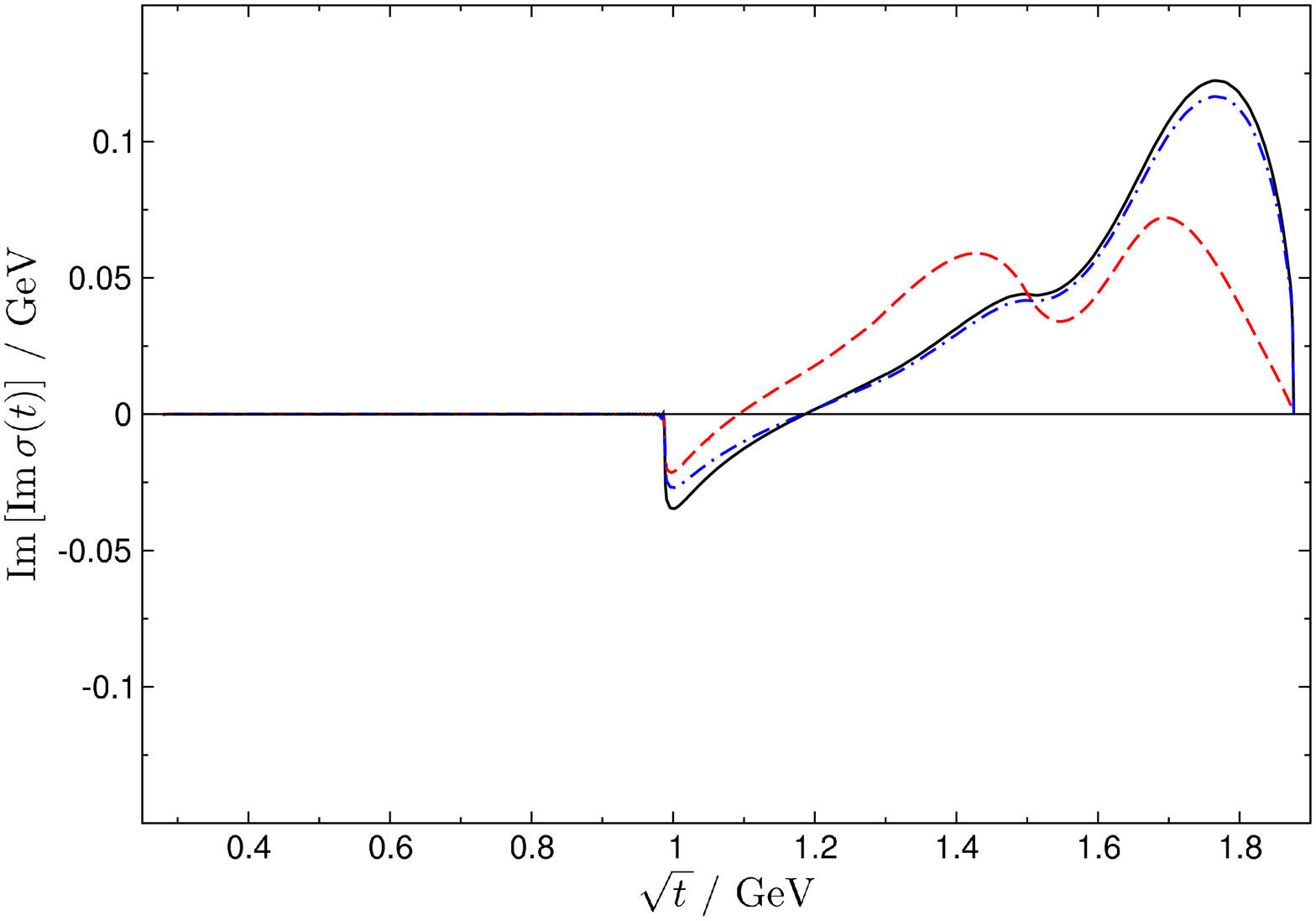}
\caption{Real and imaginary part of $\Im\sigma(t)$ for our full solution (top) and with $h^0_+(t)$ set to zero (bottom). The solid, dashed, and dot-dashed lines refer to the input RS1, RS2, and RS3 as described in Sect.~\ref{sec:pin_appl}.}
\label{fig:spectral_function}
\end{figure} 

The corresponding results using the input RS1, RS2, and RS3 as discussed in Sect.~\ref{sec:pin_appl} as well as our central solution for the scalar pion and kaon form factors are depicted in Fig.~\ref{fig:spectral_function}. We also show a variant of the spectral function where the second term in the above unitarity relation~\eqref{unitarity_sigma_term} due to $\bar K K$ intermediate states is neglected. While the impact on the real part is moderate, we see that the spectral function develops an imaginary part starting at $\tK$. In contrast, our full solution stays real as long as the input for the phases is treated in the same way in the calculation of the meson--nucleon partial waves and the scalar meson form factors. For this reason, the results for RS1 and RS3 become complex around $t_0$, while RS2 is real in the full energy range (apart from some numerical noise at the two-kaon threshold). These findings emphasize the importance of treating inelastic channels consistently in all contributions to the unitarity relation, in particular the necessity to explicitly include the intermediate states that are responsible for the inelasticities.

\begin{figure}
\centering
\includegraphics[width=0.5\textwidth]{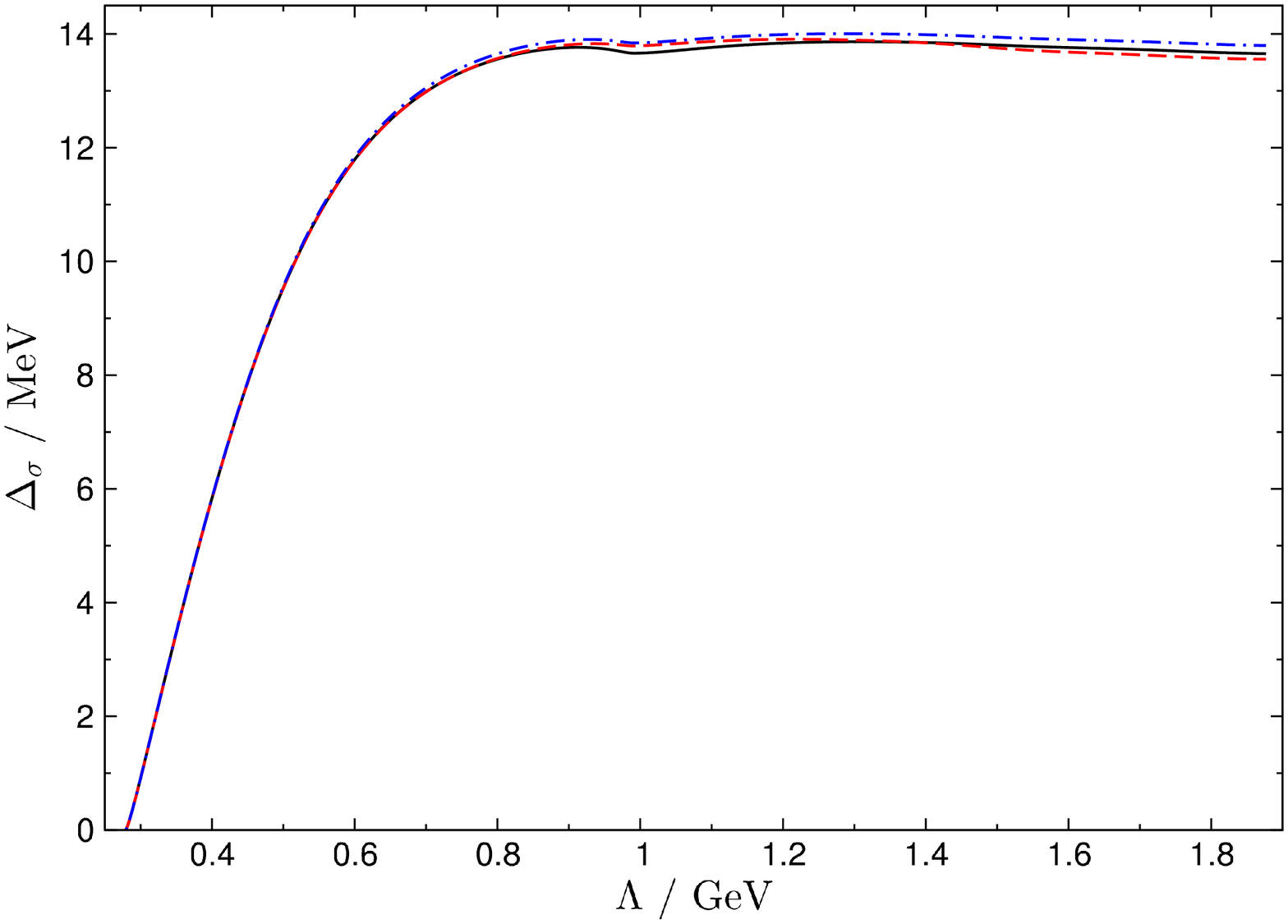}
\caption{$\Delta_\sigma$ as a function of the integral cutoff $\Lambda$. Line code as in Fig.~\ref{fig:spectral_function}.}
\label{fig:conv_disp_int}
\end{figure} 

\begin{table}
\renewcommand{\arraystretch}{1.3}
\centering
\begin{tabular}{c c c c c}
\toprule
($F^S_K(0)$, $\Lambda$) & ($0.4\,\mpi^2$, $1.3\GeV$) & ($\mpi^2/2$, $1.3\GeV$) & ($0.6\,\mpi^2$, $1.3\GeV$) & ($\mpi^2/2$, $2m$) \\
\midrule
RS1 & $13.81$ & $13.86$ & $13.91$ & $13.65$\\
RS2 & $13.75$ & $13.89$ & $14.04$ & $13.56$\\
RS3 & $13.92$ & $14.00$ & $14.09$ & $13.80$\\
\bottomrule
\end{tabular}
\renewcommand{\arraystretch}{1.0}
\caption{Results for $\Delta_\sigma$ in ${\rm MeV}$ for various combinations of $F^S_K(0)$, integral cutoff $\Lambda$, and input.}
\label{tab:DelSig}
\end{table}

The dependence of $\Delta_\sigma$ on the cutoff $\Lambda$ of the dispersive integral is shown in Fig.~\ref{fig:conv_disp_int}. We see that the dispersion relation converges quickly and the results hardly change above $1\,{\rm GeV}$. We quote the outcome for $\Lambda=\sqrt{t_0}$ and input RS2 as our final result, estimating the uncertainty by the variation induced by changing the cutoff to $2m$ and varying the input set or the assumption for $F^S_K(0)$ (cf.\ Table~\ref{tab:DelSig}). Moreover, $f^0_+(t)$ and $h^0_+(t)$ depend linearly on the $\pi N$ parameters~\eqref{subthresh_coupling}, so that the corresponding corrections for changing these parameters can be determined straightforwardly. Putting everything together, we find
\begin{align}
\label{DelSigRes}
 \Delta_\sigma&=(13.9\pm 0.3)\,{\rm MeV}\nt\\
&\qquad+ Z_1 \bigg(\frac{g^2}{4\pi}-14.28\bigg)+ Z_2\Big(d_{00}^+\,\mpi+1.46\Big)+
Z_3\Big(d_{01}^+\,\mpi^{3}-1.14\Big)+Z_4\Big(b_{00}^+\,\mpi^{3}+3.54\Big)\ec\nt\\
Z_1&=0.36\,{\rm MeV}\ec\qquad Z_2= 0.57\,{\rm MeV}\ec\qquad Z_3= 12.0\,{\rm MeV}\ec\qquad Z_4=-0.81\,{\rm MeV}\ep
\end{align}
These results are in reasonable agreement with~\cite{GLS90_FF}, and remarkably close to the $\Order(p^4)$ ChPT analysis of~\cite{BL99} (for earlier work in ChPT on $\Delta_\sigma$ see~\cite{Pagels71,Gasser82,GSS88,BKM93})
\beq
\Delta_\sigma=14.0\,{\rm MeV}+2\mpi^4\, \bar e_2\ec
\eeq
where $\bar e_2$ is an $\Order(p^4)$ low-energy constant.\footnote{In fact, $\bar e_2$ contributes to the chiral expansion of $d_{01}^+$.} The potentially largest correction in~\eqref{DelSigRes} originates from $d_{01}^+$, e.g.\ taking $d_{01}^+=1.27\,\mpi^{-3}$ from~\cite{GWU98} increases $\Delta_\sigma$ by $1.6\,{\rm MeV}$. In contrast, adjusting the coupling constant to $g^2/4\pi=13.7$~\cite{piNcoupling:short,piNcoupling:long} only leads to a correction of $-0.2\,{\rm MeV}$. Indeed, this result is not surprising, as it is $d_{01}^+$ that controls the slope of the scalar form factor of the nucleon.

\section{Conclusion}
\label{sec:concl}

We have presented a method to construct two-channel Omn\`es functions for a finite-matching-point setup, and applied this formalism to solve a coupled-channel integral equation for the $\pi\pi\to\bar N N$ and $\bar K K\to\bar N N$ $S$-waves, which can be derived from Roy--Steiner equations for $\pi N$ and $KN$ scattering.
We have also repeated the conventional two-channel Muskhelishvili--Omn\`es calculation for the scalar pion and kaon form factors, which together with the meson--nucleon partial waves determine the spectral function of the scalar form factor of the nucleon fully including the effects of $\bar K K$ intermediate states. 
Based on these results, we have updated the dispersive analysis of the correction $\Delta_\sigma$ that relates the pion--nucleon $\sigma$ term to the $\pi N$ amplitude at the Cheng--Dashen point. Throughout the calculation we have investigated the sensitivity to various pieces of the input in detail, in particular, we give our final result for $\Delta_\sigma$ as a function of $\pi N$ subthreshold parameters and the $\pi N$ coupling constant. 
Our result essentially confirms the result of~\cite{GLS90_FF}, however, using modern phase-shift input and with uncertainties due to $\bar K K$ effects much better under control.

Besides its implications for the scalar form factor of the nucleon, the present calculation of the $\pi\pi\to\bar N N$ $S$-wave in a full two-channel treatment completes the solution of the $t$-channel part of the Roy--Steiner equations derived in~\cite{RS_DHKM}, and is also a key ingredient to pin down the $\pi N$ amplitude at the Cheng--Dashen point~\cite{BL01,GLS90_FF,Hoehler}. Moreover, the formalism developed here should prove valuable for other systems requiring finite-matching-point two-channel Omn\`es solutions as well, e.g.\ for including $\bar K K$ effects into the Roy--Steiner analysis of $\gamma\gamma\to\pi\pi$~\cite{HPS}.

\subsection*{Acknowledgments}

We are grateful to Bachir Moussallam for valuable discussions and to Gilberto Colangelo for providing the results of~\cite{CCL:PWA} prior to publication.
Partial financial support by
the DFG (SFB/TR 16, ``Subnuclear Structure of Matter''),
by the project ``Study of Strongly Interacting Matter'' 
(HadronPhysics3, Grant Agreement No.~283286) 
under the 7th Framework Program of the EU, 
by the Helmholtz Association through the
Nuclear Astrophysics Virtual Institute (VH-VI-417), by the BMBF
(grant 06BN9006),
and by the Bonn--Cologne Graduate School of Physics and Astronomy
is gratefully acknowledged.

\appendix

\section{Continuity at the matching point}
\label{app:continuity}

To prove the continuity of $\ff(t)$ at $\tm$, we rewrite~\eqref{omnes_fin} in terms of a principal-value integral as
\begin{align}
 \ff(t)&=\big(\unity+i\,T(t)\Sigma(t)\big)\Del(t)+\frac{\Omega(t)}{\pi}\dashint\limits_{\tpi}^{\tm}\diff t'\frac{\Omega^{-1}(t')T(t')\Sigma(t')\Del(t')}{t'-t}+
\frac{\Omega(t)}{\pi}\int\limits_{\tm}^\infty\diff t'\frac{\Omega^{-1}(t')\Im \ff(t')}{t'-t}\ep
\end{align}
We only consider the case $t\to\tm$ from below (continuity from above can be proven in a similar way). In this limit, the whole mass of the integral is concentrated at $\tm$. Using~\eqref{def_below}, \eqref{def_above}, \eqref{integrals}, and~\eqref{xij}, we find
\begin{align}
\label{cont_cond}
 \ff(\tm)&=\big(\unity+i\,T(\tm)\Sigma(\tm)\big)\Del(\tm)+\frac{e^{-i\pi x}}{\det\bar\Omega(\tm)}\begin{pmatrix} I_{\rm I} & I_{\rm II}\\ I_{\rm III} & I_{\rm IV}  \end{pmatrix} T(\tm)\Sigma(\tm)\Del(\tm)\nt\\
&\qquad- \frac{1}{\det\bar\Omega(\tm)}\begin{pmatrix} \tilde I_{\rm I} & \tilde I_{\rm II}\\ \tilde I_{\rm III} & \tilde I_{\rm IV}  \end{pmatrix} \Im\ff(\tm)\ec                                                                                      
\end{align}
with
\begin{align}
\label{def_I}
 I_{\rm I}&=-\Big(\bar\Omega_{11}\bar\Omega_{22}\cot\pi x_{11}-\bar\Omega_{12}\bar\Omega_{21}\cot\pi x_{12}\Big)e^{i\psi}\ec
&I_{\rm II}&=\bar\Omega_{11}\bar\Omega_{12}\Big(\cot\pi x_{11}-\cot\pi x_{12}\Big)e^{i\psi}\ec\nt\\
I_{\rm III}&=-\bar\Omega_{21}\bar\Omega_{22}\Big(\cot\pi x_{11}-\cot\pi x_{12}\Big)e^{i\psi}\ec
&I_{\rm IV}&=-\Big(\bar\Omega_{11}\bar\Omega_{22}\cot\pi x_{12}-\bar\Omega_{12}\bar\Omega_{21}\cot\pi x_{11}\Big)e^{i\psi}\ec\nt\\
\tilde I_{\rm I}&=-\bar\Omega_{11}\bar\Omega_{22}\frac{e^{i\delta_{11}}}{\sin\pi x_{11}}+\bar\Omega_{12}\bar\Omega_{21}\frac{e^{i\delta_{12}}}{\sin\pi x_{12}}\ec
&\tilde I_{\rm II}&=\bar\Omega_{11}\bar\Omega_{12}\bigg\{\frac{e^{i\delta_{11}}}{\sin\pi x_{11}}-\frac{e^{i\delta_{12}}}{\sin\pi x_{12}}\bigg\}\ec\nt\\
\tilde I_{\rm III}&=-\bar\Omega_{21}\bar\Omega_{22}\bigg\{\frac{e^{i\delta_{11}}}{\sin\pi x_{11}}-\frac{e^{i\delta_{12}}}{\sin\pi x_{12}}\bigg\}\ec
&\tilde I_{\rm IV}&=-\bar\Omega_{11}\bar\Omega_{22}\frac{e^{i\delta_{12}}}{\sin\pi x_{12}}+\bar\Omega_{12}\bar\Omega_{21}\frac{e^{i\delta_{11}}}{\sin\pi x_{11}}\ep
\end{align}
For the continuity condition~\eqref{cont_cond} to be fulfilled, in particular all terms depending on $\Del(\tm)$ must cancel amongst themselves. Putting the coefficients of $\Delta_i(\tm)$ in each component $f_i$ to zero yields four constraints on $I_{\rm I}$--$I_{\rm IV}$, which can be inverted to obtain
\begin{align}
\label{sol_I}
 I_{\rm I}&=i\det \bar\Omega\, e^{i\pi x}\frac{A-e^{2i\pi x}A^*}{A+e^{2i\pi x}A^*}\ec\qquad 
I_{\rm IV}=i\det \bar\Omega\, e^{i\pi x}\frac{B-e^{2i\pi x}B^*}{A+e^{2i\pi x}A^*}\ec\nt\\
I_{\rm II}&=\frac{\sigma_{\tm}^K}{\sigma_{\tm}^\pi}I_{\rm III}=4\det \bar\Omega\, e^{2i\pi x}\frac{|g|\sigma_{\tm}^K}{A+e^{2i\pi x}A^*}\ec
\end{align}
where
\beq
A=\eta e^{2i\pi(x-y)}-1\ec\qquad B=\eta e^{2i\pi y}-1\ep
\eeq
Similar considerations apply to $\tilde I_{\rm I}$--$\tilde I_{\rm IV}$. By means of $\Im \ff=T^*\Sigma \ff$, the remaining pieces of~\eqref{cont_cond} can be expressed as linear combinations of $f_1$ and $f_2$, which altogether again amounts to four constraints. The solutions are
\beq
\label{sol_tilde_I}
 \tilde I_{\rm I}=-2i\,\det \bar\Omega\, e^{2i\pi x}\frac{A^*}{A+e^{2i\pi x}A^*}\ec\qquad\!\! 
\tilde I_{\rm IV}=-2i\,\det \bar\Omega\, e^{2i\pi x}\frac{B^*}{A+e^{2i\pi x}A^*}\ec\qquad\!\!
\tilde I_{\rm II}=\frac{\sigma_{\tm}^K}{\sigma_{\tm}^\pi}\tilde I_{\rm III}=I_{\rm II}e^{-i\pi x}\ep
\eeq
Based on~\eqref{sinx11}
it is straightforward to show that~\eqref{sol_I} and~\eqref{sol_tilde_I} hold as long as the following relations are fulfilled
\begin{align}
\label{cont_conditions}
 \sqrt{1-z^2}\big(\bar\Omega_{11}\bar\Omega_{22}+\bar\Omega_{12}\bar\Omega_{21}\big)&=\eta\sin\pi(2y-x) \det\bar\Omega\ec\nt\\
\sqrt{1-z^2}\bar\Omega_{11}\bar\Omega_{12}&=-|g|\sigma_{\tm}^K\det\bar\Omega\ec\qquad\qquad \sqrt{1-z^2}\bar\Omega_{21}\bar\Omega_{22}=|g|\sigma_{\tm}^\pi\det\bar\Omega\ep
\end{align}
In this way, the properties~\eqref{ratio_Omij} of the Omn\`es matrix finally ensure continuity at the matching point.

\section{Conventions for kaon--nucleon scattering}
\label{app:KN}

In general, we follow the notation and conventions for $\pi N$ scattering of~\cite{RS_DHKM}. In this appendix, we briefly collect the additional formalism necessary to arrive at a closed system for $\pi N$ and $K N$ scattering.

We define the kaon states
\beq
|K^+\rangle=\left|\frac{1}{2},\frac{1}{2}\right\rangle\ec\qquad |K^0\rangle=\left|\frac{1}{2},-\frac{1}{2}\right\rangle\ec \qquad 
|\bar K^0\rangle=\left|\frac{1}{2},\frac{1}{2}\right\rangle\ec\qquad |K^-\rangle=\left|\frac{1}{2},-\frac{1}{2}\right\rangle\ec
\eeq
with crossing properties
\beq
C|K^+\rangle=-|K^-\rangle\ec \qquad C|K^0\rangle=|\bar K^0\rangle\ec
\eeq
and choose the isospin decomposition of the amplitude as
\beq
T=T^+-\boldsymbol{\tau}_N\cdot \boldsymbol{\tau}_K T^-=T^+-2\Big(I_s(I_s+1)-\frac{3}{2}\Big)T^-\ec
\eeq
which leads to
\begin{align}
\label{cJKN}
T^{I_s=0}&=T^++3T^-\ec &T^{I_s=1}&=T^+-T^-\ec\nt\\
T^{I_u=0}&=T^+-3T^-\ec &T^{I_u=1}&=T^++T^-\ec\nt\\
T^{I_t=0}&=2T^+\ec &T^{I_t=1}&=2T^-\ep 
\end{align}
In these equations $I_s$, $I_t$, and $I_u$ denote $s$-, $t$-, and $u$-channel isospin, respectively, and $\boldsymbol{\tau}_N$ and $\boldsymbol{\tau}_K$ are the Pauli matrices associated with the nucleon and kaon isospin operators.
In particular, we have
\beq
T^{K^\pm p}=T^{K^\pm p\to K^\pm p}=T^+\mp T^-\ep
\eeq
In complete analogy to $\pi N$ scattering, the amplitude for the process $K(q)+N(p)\to K(q')+N(p')$ can be decomposed as
\beq
T^I=\bar u (p')\bigg\{A^I+\frac{\slashed q'+\slashed q}{2}B^I\bigg\} u(p)\ec\qquad I\in\{+,-\} \ep
\eeq
The Born-term contributions due to hyperon pole diagrams are\footnote{We use $m_\Lambda=1.116\,{\rm GeV}$, $m_\Sigma=1.193\,{\rm GeV}$~\cite{PDG}, and $g_{KN\Lambda}^2/4\pi=15.55$, $g_{KN\Sigma}^2/4\pi=0.576$~\cite{Speth89} for the masses and couplings of the hyperons, respectively.} 
\begin{align}
A^{K^+p}&=\sum_{Y={\Lambda,\Sigma}}g_{KNY}^2\frac{m_Y-m}{u-m_Y^2}\ec\qquad B^{K^+p}=\sum_{Y={\Lambda,\Sigma}}\frac{g_{KNY}^2}{u-m_Y^2}\ec\nt\\
A^{K^-p}&=\sum_{Y={\Lambda,\Sigma}}g_{KNY}^2\frac{m_Y-m}{s-m_Y^2}\ec\qquad B^{K^-p}=\sum_{Y={\Lambda,\Sigma}}\frac{g_{KNY}^2}{s-m_Y^2}\ec
\end{align}
and hence
\beq
 A^{\pm}=\sum_{Y={\Lambda,\Sigma}}g_{KNY}^2\frac{m_Y-m}{2}\bigg(\frac{1}{s-m_Y^2}\pm\frac{1}{u-m_Y^2}\bigg)\ec\qquad
B^{\pm}=-\sum_{Y={\Lambda,\Sigma}}\frac{g_{KNY}^2}{2}\bigg(\frac{1}{s-m_Y^2}\mp\frac{1}{u-m_Y^2}\bigg)\ep
\eeq
The $\bar K K\to\bar N N$ partial waves can be obtained from the invariant amplitudes by means of the projection formula (with the $t$-channel scattering angle $z_t=\cos\theta_t^{KN}$)
\begin{align}
\label{KN_tchannel_proj}
 h_+^{J,I_t=0,1}(t)&=-\frac{1}{8\pi}\int\limits^1_{-1}\diff z_tP_J(z_t)\bigg\{\frac{p_t^2}{(p_tk_t)^J}A^{\pm}(t,z_t)-\frac{m}{(p_tk_t)^{J-1}}z_tB^{\pm}(t,z_t)\bigg\}\ec\nt\\
 h_-^{J,I_t=0,1}(t)&=\frac{1}{8\pi}\frac{\sqrt{J(J+1)}}{2J+1}\frac{1}{(p_tk_t)^{J-1}}\int\limits^1_{-1}\diff z_t\big(P_{J-1}(z_t)-P_{J+1}(z_t)\big)B^{\pm}(t,z_t)\ep
\end{align}
The main difference to the $\pi N$ $t$-channel partial-wave projection~\cite{FrazerFulco:tPW} originates from the fact that due to the lack of Bose symmetry in the $\bar K K$ system a partial wave with given angular momentum $J$ couples to both $I_t=0$ and $I_t=1$ (corresponding to $+$ and $-$ on the right-hand side of~\eqref{KN_tchannel_proj}).
In the following, we are only interested in the combination where even/odd $J$ corresponds to $I_t=0,1$, respectively, since only these partial waves can occur as intermediate states in $\pi\pi\to\bar N N$, and will therefore suppress the isospin index. In these conventions, the Born terms are given by
\begin{align}
\label{hyperon_pole}
 h_+^J(t)&=\sum_{Y={\Lambda,\Sigma}}\frac{g_{KNY}^2}{8\pi}\frac{1}{(p_tk_t)^J}\bigg\{\Big(\frac{p_t}{k_t}(m_Y-m)+m\tilde y\Big)Q_J(\tilde y)-m\delta_{J0}\bigg\}\ec\nt\\
 h_-^J(t)&=\sum_{Y={\Lambda,\Sigma}}\frac{g_{KNY}^2}{8\pi}\frac{\sqrt{J(J+1)}}{2J+1}\frac{1}{(p_tk_t)^J}\big(Q_{J-1}(\tilde y)-Q_{J+1}(\tilde y)\big)\ec
\end{align}
with
\beq
\tilde y=\frac{t-2M_K^2+2(m_Y^2-m^2)}{4p_tk_t}\ep
\eeq

The partial waves in the partial-wave expansion of the $\bar K K$ scattering amplitude
\beq
R^{I_t}(s,t)=16\pi\sum\limits_{J=0}^\infty(2J+1)r^{I_t}_J(t)P_J(\cos\theta_t^{\bar K K})
\eeq 
obey the unitarity relation
\begin{align}
\Im r^{I_t}_J(t)&=\sigma^\pi_t\Big|(q_tk_t)^Jg_J^{I_t}(t)\Big|^2\,\theta\big(t-\tpi\big)+\sigma^K_t\Big|r_J^{I_t}(t)\Big|^2\,\theta\big(t-\tK\big)\nt\\
&\qquad+\frac{t\sigma^N_t}{8k_t^2}\frac{1}{(c^{KN}_J)^2}\bigg\{\Big|H_+^J(t)\Big|^2+\Big|H_-^J(t)\Big|^2\bigg\}\theta\big(t-\tN\big)\ec
\end{align}
where $g_J^{I_t}(t)$ denotes the $\pi\pi\to\bar K K$ partial waves, the $\bar K K\to\bar N N$ amplitudes fulfill
\beq
H_+^J(t)=\frac{k_t}{p_t}(p_tk_t)^J\frac{2}{\sqrt{t}}h_+^J(t)\ec\qquad
 H_-^J(t)=\frac{k_t}{p_t}(p_tk_t)^Jh_-^J(t)\ec
\eeq
and $c^{KN}_J=1/2$ is an isospin factor that emerges from the conversion between the $I_t=0,1$ and the $I=\pm$ basis (cf.~\eqref{cJKN}). 
With the $S$-matrix elements
\beq
\big[S^{I_t}_J(t)\big]_{\bar KK\to\bar KK}=1+i\frac{4k_t}{\sqrt{t}}r^{I_t}_J(t)\,\theta\big(t-\tK\big)\ec\qquad
\big[S^J_\pm(t)\big]^{I_t}_{\bar K K\to\bar NN}=\frac{i}{c_J^{KN}}\sqrt{\frac{p_t}{k_t}}H^J_\pm(t)\,\theta\big(t-\tN\big)\ec
\eeq
we finally obtain the unitarity relation
\beq
\Im h^J_{\pm}(t)=\sigma^K_t\big(r^{I_t}_J(t)\big)^*h^J_{\pm}(t)\,\theta\big(t-\tK\big)
+\frac{c_J^{KN}}{\sqrt{2}\,c_J}\,\sigma^\pi_tq_t^{2J}\big(g^{I_t}_J(t)\big)^*f_\pm^J(t)\,\theta\big(t-\tpi\big)\ec
\eeq
with $\pi\pi\to\bar N N$ partial waves $f_\pm^J(t)$ and isospin factors $c_J$ for $\pi N$ scattering (see~\cite{RS_DHKM} for precise definitions).

\end{document}